\documentclass[
 prl,twocolumn,
 amsmath,amssymb,
 aps,nofootinbib,superscriptaddress,
]{revtex4-2}

\usepackage{amsthm}
\usepackage{graphicx}
\usepackage{color}
\usepackage{dcolumn}
\usepackage{bm}
\usepackage{mathtools}      % \coloneqq
\usepackage{mathrsfs}       % for operator symbol
\usepackage{braket}
\usepackage{float}
\usepackage[table]{xcolor}

\newcommand{\cD}{\ensuremath{\mathcal{D}}}
\newcommand{\cE}{\ensuremath{\mathcal{E}}}
\newcommand{\cF}{\ensuremath{\mathcal{F}}}

\newcommand{\cC}{\ensuremath{\mathcal{C}}}

\newcommand{\cQ}{\ensuremath{\mathcal{Q}}}

\newcommand{\cL}{\ensuremath{\mathcal{L}}}

\newcommand{\bth}{\ensuremath{\boldsymbol{\theta}}}
\newcommand{\bzet}{\ensuremath{\boldsymbol{\zeta}}}

\newcommand{\bE}{\ensuremath{\boldsymbol{E}}}

\DeclareMathOperator{\tr}{\textup{Tr}}

\DeclareMathAlphabet{\mymathbb}{U}{BOONDOX-ds}{m}{n}

\usepackage{hyperref}
\hypersetup{
    colorlinks = true,
    linkcolor = {blue},
    citecolor = {blue},
    urlcolor = {blue},
}

\begin{document}
\title{Variational Denoising for Variational Quantum Eigensolver}

\author{Quoc Hoan Tran}
\email{tran.quochoan@fujitsu.com}
\affiliation{Quantum Laboratory, Fujitsu Research, Fujitsu Limited, Kawasaki, Kanagawa 211-8588, Japan}

\author{Shinji Kikuchi}
\affiliation{Quantum Laboratory, Fujitsu Research, Fujitsu Limited, Kawasaki, Kanagawa 211-8588, Japan}

\author{Hirotaka Oshima}
\affiliation{Quantum Laboratory, Fujitsu Research, Fujitsu Limited, Kawasaki, Kanagawa 211-8588, Japan}

\date{\today}

\begin{abstract}

The variational quantum eigensolver (VQE) is a hybrid algorithm that has the potential to provide a quantum advantage in practical chemistry problems that are currently intractable on classical computers. VQE trains parameterized quantum circuits using a classical optimizer to approximate the eigenvalues and eigenstates of a given Hamiltonian. However, VQE faces challenges in task-specific design and machine-specific architecture, particularly when running on noisy quantum devices. This can have a negative impact on its trainability, accuracy, and efficiency, resulting in noisy quantum data. We propose variational denoising, an unsupervised learning method that employs a parameterized quantum neural network to improve the solution of VQE by learning from noisy VQE outputs. Our approach can significantly decrease energy estimation errors and increase fidelities with ground states compared to noisy input data for the $\text{H}_2$, LiH, and $\text{BeH}_2$ molecular Hamiltonians, and the transverse field Ising model. Surprisingly, it only requires noisy data for training. Variational denoising can be integrated into quantum hardware, increasing its versatility as an end-to-end quantum processing for quantum data.

\end{abstract}

\pacs{Valid PACS appear here}

\maketitle
\textit{Introduction.---} The term ``quantum advantage'' originates from the potential of quantum computers to solve practical problems much faster than even the most powerful classical computer. 
While fault-tolerant quantum computers are expected to provide a quantum advantage, noisy effects in relevant quantum hardware necessitate a large number of qubits for useful and reliable computation.
Nevertheless, the noisy intermediate-scale quantum (NISQ) devices~\cite{preskill:2018:NISQ} continue to excite the research community by promising to outperform classical computers in certain mathematical tasks~\cite{neill:2018:science:supremacy,arute:2019:nature:supremacy,zhong:2020:science:supremacy,zhong:2021:prl:supremacy}.
Following the success of classical neural networks (NNs) with noise-robust trainability for various practical machine learning problems, there has recently been renewed optimism that quantum NNs on NISQ computers can lead to quantum advantage.
Although quantum speedup has been demonstrated in impractical supervised learning tasks using artificial and engineered datasets~\cite{liu:2020:rigorous}, efforts are being made to identify complex distributions where classical NNs are inherently inefficient but quantum NNs can excel~\cite{gao:2022:prx:gen}.

%Indeed, the quantum speedup has been demonstrated in an impractical supervised learning problem using an artificial and engineered data set~\cite{liu:2020:rigorous}.
%There has been progress in identifying complex distributions that classical NNs are provably inefficient in expressing but quantum NNs~\cite{gao:2022:prx:gen}.

Training parameterized quantum circuits (PQCs) with variational quantum algorithms (VQAs) is an important factor in NISQ devices~\cite{cezero:2021:natrev:VQA}. In VQA, an optimization problem is designed with a loss function computed from the expectation values of observables in a PQC. Classical computers are used to minimize this loss function by adjusting the circuit's parameters. However, the trainability of PQCs can be hampered by factors such as the instability of stochastic optimizers in complex loss landscapes, quantum errors, and systematic errors in experimental setups. The well-known barren plateau phenomenon~\cite{clean:2018:natcom:barren}, where the loss landscape is effectively flat, can occur under certain conditions such as deep quantum circuits, nonlocality of the observable defining the loss~\cite{cerezo:2021:natcom:barren}, and excessive quantum entanglement~\cite{marrero:2021:prxquant:barren}.
Even when using shallow and local quantum models with local cost functions to avoid the barren plateau, poor local optimal concentrations far from the global optimum can lead to  untrainability~\cite{ansachuetz:2022:natcom:VQA,bittel:2021:prl:VQANP}.
Furthermore, factors such as quantum noise, task-specific circuit design, and hyperparameter settings, can have a significant impact on training performance.

Two representative methods in VQAs are quantum approximate optimization algorithms (QAOAs)~\cite{farhi:2014:QAOA} and variational quantum eigensolvers (VQEs)~\cite{peruzzo:2014:VQE}. 
%QAOAs are proposed for combinatorial optimization problems, such as finding the maximum cut or the largest complete subgraph of a graph. 
%Recent research shows that under certain regimes, QAOA can outperform certain classical algorithms when applied to an energy minimization problem in spin models~\cite{farhi:2022:QAOA:SKmodel}.
VQEs are designed for applications in quantum chemistry and material science to determine the lowest energy level and  ground state of a Hamiltonian. Since identifying solutions to such problems on classical computers incurs a computational cost that exponentially grows with the system's size, VQEs hold the promise of demonstrating the quantum advantage~\cite{sokolov:2020:qUCC}. Despite the successful implementation on small-scale problems~\cite{kandala:2017:nature:VQE}, large-scale implementations of VQE remain a challenge~\cite{jules:2022:physrep:vqe:review}.

%overview_v6.pdf
\begin{figure}
		\includegraphics[width=8.7cm]{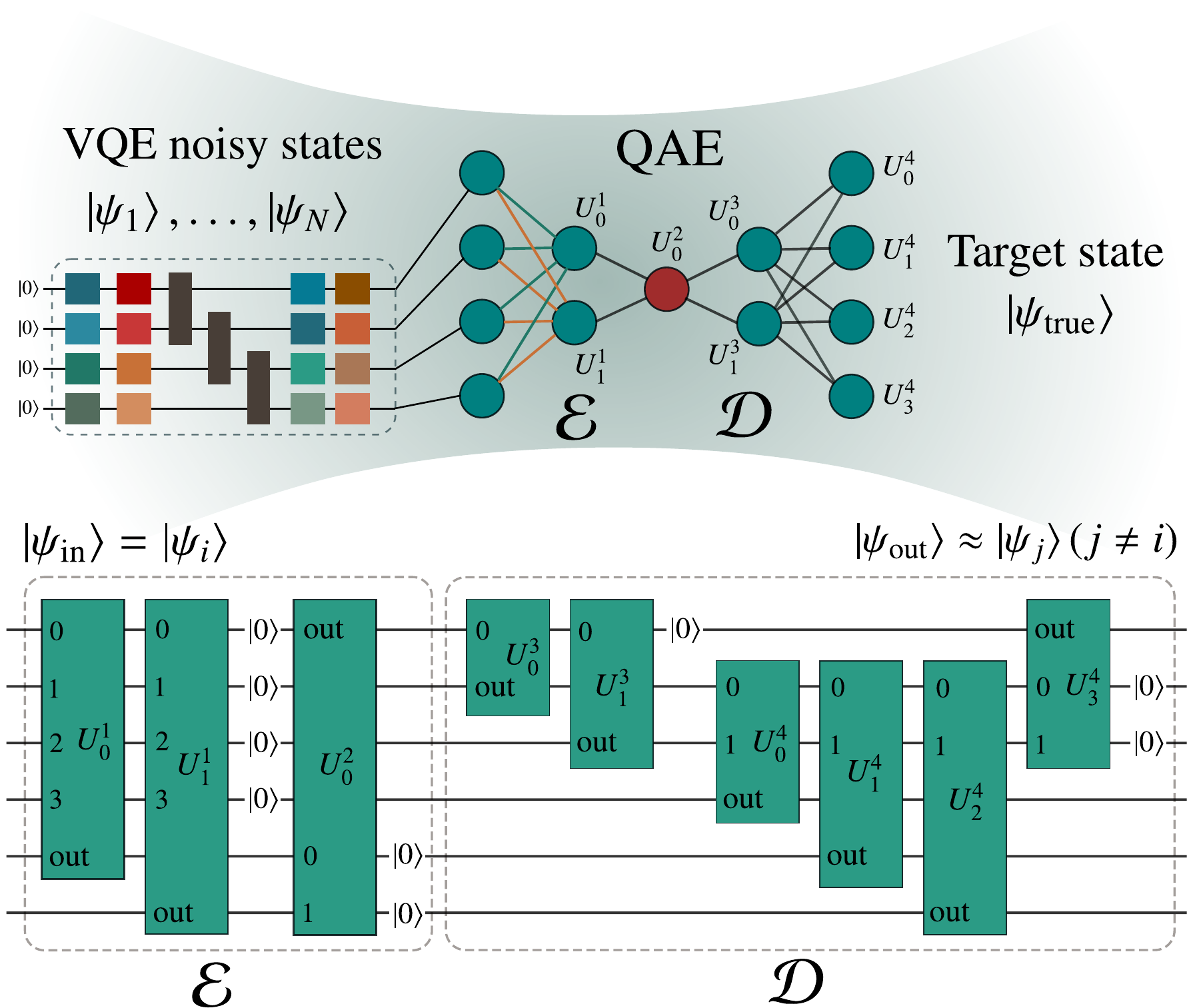}
		\protect\caption{Variational denoising by QAE.  Quantum states prepared by VQE circuits are considered ``noisy" states. QAE trains encoding and decoding PQCs to compress each noisy state into a subsystem so that the state recovered from the reduced states is as close as possible to other noisy states. QAE is generalized from the classical feedforward NN where each neuron corresponds to a qubit while unitary circuits connect neurons in subsequent coupled layers (upper panel). The number of neurons is reduced in the hidden layer to compress the input state. The lower panel describes an example of circuit-based implementation for a QAE[4,2,1,2,4].
		\label{fig:qae:overview}}
\end{figure}

VQE generates quantum states $\ket{\psi({\bzet})}=V(\bzet)\ket{\psi}$ by applying unitary $V(\bzet)$ of parameter $\bzet$ to some arbitrary starting state $\ket{\psi}$. 
Given a Hermitian operator $\hat{H}$, VQE provides an estimation on the ground state $\ket{\psi_{\text{min}}}$ which corresponds to an unknown minimum eigenvalue (lowest energy) $E_{\text{min}}$ of $\hat{H}$.
The energy estimation $E_{\bzet}$ of VQE is bounded as $E_{\text{min}} \leq E_{\bzet} = \bra{\psi(\bzet)} \hat{H} \ket{\psi(\bzet)}$,
which is then optimized by classical optimizers to obtain the minimum value. 
%The goal is to estimate the state $\ket{\psi({\bzet})}$ with the highest overlap with the ground state $\ket{\psi_{\text{min}}}$. 
However, the estimation of $E_{\bzet}$ can be a barrier to achieving near-term practical quantum advantages, since the number of quantum measurements required can be prohibitive~\cite{gonthier:2022:VQEblock}. This can prevent VQE from being run for a large number of iterations~\cite{ansachuetz:2022:natcom:VQA}. Furthermore, stochasticity in the optimizer and noise in quantum devices can result in different final $\ket{\psi({\bzet})}$ as $\ket{\psi_1},\ldots,\ket{\psi_N}$. These states are considered noisy data providing only partial information about the target ground state.

Here, we propose a method for learning the target state from noisy VQE data.
This idea can be realized using the quantum extension of the autoencoder framework~\cite{good:2016:DL}, in which feedforward NNs are trained to reproduce the input at the output layer.
In the autoencoder, the number of nodes in the hidden layer is typically less than the number of nodes in the input and output layers, indicating that some essential information is retained for the reconstruction. 
The quantum autoencoder (QAE) is described in Ref.~\cite{romero:2017:QAE} with applications in compressing and denoising a specific quantum state~\cite{bondarenko:2020:prl:QAE,apache:2020:QAE,locher:2022:QAE,pazem:2023:brainbox}. 
Our proposal enhances the practical applicability of QAE by aiming for a hardware-efficient approach that learns from noisy quantum data. It serves as a denoising method, not only for specific states but also focuses on improving the robustness and efficiency of VQE.

\textit{Quantum Autoencoder (QAE).---} QAE generates an encoder $\cE$ to map the $n$-qubits input state  to a $k$-qubits ($k<n$) reduced state, and a decoder $\cD$ to reconstruct the input state from the reduced state.
$\cE$ and $\cD$ are implemented using PQCs with parameters $\bth_e$ and $\bth_d$, respectively.
They are trained to maximize the expectation fidelity $\cF$ of the input state $\rho^{\text{in}}$ with the output of QAE over the input data distribution $\cC$ 
\begin{align}\label{eqn:expect}
    \max_{\bth_e,\bth_c}\bE_{\rho^{\text{in}} \sim \cC}\left[\cF(\rho^{\text{in}}, \cD(\cE(\rho^{\text{in}}))\right].
\end{align}
In practice, when considering the pure states $\ket{\psi_1},\ldots,\ket{\psi_N}$ as training data sampled from $\cC$, Eq.~\eqref{eqn:expect} can reduce to maximize the empirical cost 
\begin{align}\label{eqn:empirical}
    \cL(\bth_e, \bth_c) = \dfrac{1}{N}\sum_{i=1}^N\cF(\ket{\psi_i}, \cD(\cE(\ket{\psi_i}))).
\end{align}
Here, the fidelity between a pure state $\ket{\psi}$ and a density matrix $\rho$ is defined as $\cF(\ket{\psi}, \rho)=\braket{\psi|\rho|\psi}$.
This quantity can be implemented using the SWAP test, which is presented in our training circuit in~\cite{supp}.
%The QAE is a quantum circuit that uses VQA to compress input quantum data into a smaller latent space. %Given the training data $\{\rho_1, \ldots, \rho_N\}$, .

The upper panel of Fig.~\ref{fig:qae:overview} shows an example of a QAE based on a dissipative quantum NN~\cite{beer:2020:DQNN,sharma:2022:dissipative}.
Here, a dissipative quantum NN generalizes the classical feedforward NN with each neuron representing a qubit and unitary circuits connecting neurons in subsequent coupled layers.
Qubits in each layer are dissipative after the forward process in the next layer, allowing the reduced space for QAE to be constructed.
Consider the QAE with structure $[m_1, m_2, \ldots, m_{M}]$ ($M\geq 2$), for neuron $j$th in the layer $i+1$, we denote $U_j^i$ as the parameterized unitary acting on its own qubit and the neurons on the preceding layer.
%These unitary matrices are subject to optimization.
The unitary between layers $i$ and $i+1$ are denoted by $U^i=\prod_j U_j^i$.
We denote $\beta_i$ as the state at the $i$th layer with $m_i$ neurons. Therefore, the transformation between $\beta_{i}$ and $\beta_{i+1}$ can be expressed via the channel
\begin{align}
    \beta_{i+1} = \cQ_i(\beta_{i}) = \tr_i\left[ U^i(\beta_i \otimes (\ket{0}\bra{0})^{\otimes m_{i+1}})(U^{i})^{\dagger} \right]. 
\end{align}
Here, $\cQ_i$ maps an $m_i$-qubit state to an $(m_i + m_{i+1})$-qubit state before reducing it to an $m_{i+1}$-qubit state by tracing out qubits in layer $i$.
For example, in the QAE[4,2,1,2,4] in Fig.~\ref{fig:qae:overview}, the encoder $\cE=\cQ_1\cQ_2$ comprises two channels: $\cQ_1$ extending $4$-qubit states (layer 1) into a $6$-qubit state and then reducing it to a $2$-qubit state (layer 2), and $\cQ_2$ extending $2$-qubit states into a $3$-qubit state and then reducing it to a $1$-qubit state (layer 3).
The decoder $\cD=\cQ_3\cQ_4$ inversely works on this $1$-qubit state to reconstruct the final $4$-qubit state.
QAE can be implemented in the quantum circuit with the number of qubits is $\max\{(m_i+m_{i+1})\}$ (lower panel in Fig.~\ref{fig:qae:overview}).
%which is larger than the number of original qubits for the input state. Another simple implementation for QAE~\cite{romero:2017:QAE} is illustrated in the lower panel of Fig.~\ref{fig:qae:overview}, where $\cE$ and $\cD$ are the sequences of quantum gates acting on the original qubits.
% Between $\cE$ and $\cD$, a qubit-reset process on $n-k$ qubits is performed to obtain reduced quantum states before going to the reconstruction.

Implementations of QAE in Refs.~\cite{bondarenko:2020:prl:QAE,apache:2020:QAE,pazem:2023:brainbox} optimize over the space of unitary operators, limiting the scale of experimentation.
We present an ansatz-based implementation of QAE that is hardware efficient and scales well with the number of qubits as
\begin{align}
    U_j^i(\bth) = \prod_{l=1}^L \left[R^{(l)}_{\text{loc}}(\bth_l)V_{\text{ent}}\right] R^{(L+1)}_{\text{loc}}(\bth_{L+1}).
\end{align}
Here, each unitary $U_j^i$ consists of $L$ blocks and final single qubit rotation on every qubit. Each block comprises of local single qubit rotations $R^{(l)}_{\text{loc}}(\bth_l)=\otimes_k R(\theta_{k,l})$ as well as two-qubit entangling gates. 
In the following experiments, we use the ansatz RY\_CZ, where the single-qubit rotation gate $R$ is the parametric rotation gate $R_Y(\theta) = e^{-iY\theta/2}$ of Pauli Y matrix, and the entangler $V_{\text{ent}}$ is composed of controlled-phase gates $CZ(k,k+1)$ placed in circular with indexes $k$ of qubits.
%Our implementation is hardware efficient and can reduce the number of parameters for training. 
The number of parameters only scales linearly with the number of qubits and ansatz blocks.
In comparison to Refs.~\cite{bondarenko:2020:prl:QAE,apache:2020:QAE}, this is significantly efficient in terms of training resources but still requires further improvement when scaling up.

\textit{Denoising QAE.---}Since QAE discards irrelevant information from the input in the compressed state, it can be used for denoising quantum data. This concept is inspired by the classical denoising autoencoder~\cite{vincent:2008:DAE}, which is trained to reconstruct a clean input from a corrupted version. In Ref.~\cite{locher:2022:QAE}, noise-free logical states are used as target states to train the QAE to remove noise from states in a predefined codespace. However, because of the no-cloning theorem, preparing copies of the target state as a reference state in quantum training is often challenging.
We can overcome this problem by using pairs of input and target states from the same noisy data source such as a quantum state with bitflip and dephasing errors, and small random unitary transformations~\cite{bondarenko:2020:prl:QAE}. However, current QAE denoising applications are limited on specific families of quantum states, such as GHZ, W, Dicke, and cluster states~\cite{bondarenko:2020:prl:QAE}. We show how QAE can be applied to VQEs where quantum noise and stochastic behavior of optimizers can result in noisy final states. 
%We train the QAE on paired states to extract essential information from the noiseless final state.
%Even the paired states will be different due to different noise realization, the QAE can extract the common essential features between them, which are used to reconstruct the clean quantum data.

Let us consider a final quantum state $\rho$ in a VQE running with a specific setting, such as initialization, the number of iterations for the optimizer, and a quantum noise condition. We can assume that $\rho$ is generated by a distribution $\cC_{\rho^*}$ based on the target state $\rho^{*}$. For example, $\rho^*$ is the ground state, and $\cC_{\rho^*}$ is the process that yields the final VQE solution after a fixed number of iterations in a stochastic optimizer.
The training procedure attempts to maximize the empirical loss
\begin{align}\label{eqn:emp:denoise}
    \cL(\bth_e, \bth_c) = \dfrac{1}{N}\sum_{i=1}^N\cF(\rho^A_i, \cD(\cE(\rho^B_i)),
\end{align}
given $N$ paired training states $(\rho^A_i, \rho^B_i)$ where $\rho^A_i, \rho^B_i$ are drawn from $\cC_{\rho^*}$.
We consider $\cC_{\rho^*}$ to be a noisy data source, which can generate different realizations of the states.
If these states share essential features of $\rho^*$, training QAE can output the state with a high overlap with the target state.
In our optimization, we combine the simultaneous perturbation stochastic approximation (SPSA) method~\cite{spall:1999:SPSA} with AMSGrad method~\cite{reddi:2018:AmsGrad}. 
SPSA is efficient because the number of measurements is independent of the total number of training parameters~\cite{supp}.

\textit{Results.---}We apply QAE to denoising states generated by the VQE algorithm.
In the following experiments, we consider the QAE circuit in a well-controlled platform without the effect of quantum noise but discuss this impact in~\cite{supp}.
The noisy data can be considered as the state generated from a noisy quantum channel applied to the VQE circuits (the first experiment) or the results of the early stopping in the optimization process of VQE (the second and third experiment). In~\cite{supp}, we also consider noisy data generated from the combination of quantum noise and early stopping optimization or from different ansatzes of the VQE circuit.
%Our method can improve the noisy solutions to a state with a high overlap with the target ground state.
We perform numerical experiments using Qiskit~\cite{Qiskit,ibm:challenge:2021} simulation for $\text{H}_2$ and LiH molecules, in which the molecular Hamiltonians presented in the Slater-type orbital (STO-3G) basis are mapped to qubit Hamiltonians with 2 (for $\text{H}_2$) and 4 (for LiH) qubits, respectively~\cite{bravyi:2002:fermionic,seeley:2012:parity}. 
For VQE circuits, we construct hardware-efficient circuits comprising single-qubit operations spanned by SU(2) and two-qubit controlled-X entangling gates~\cite{barkoutsos:2018:twolocal}. Here, we use rotation operators of Pauli Y and Z as single qubit gates~\cite{supp}.

In the first experiment, we consider process $\cC_{\rho^*}$ as a bitflip or depolarizing noise channel applied at the end of VQE circuits after performing SPSA optimization of 250 iterations for $\text{H}_2$ and 1000 iterations for LiH. In the bitflip channel, each qubit is applied to the Pauli X gate with a probability of $p_b$. In the depolarization channel, each qubit is applied to one of the gates \{I, Pauli X, Pauli Y, Pauli Z\} with respective probabilities $\{1-\frac{3p_d}{4},\frac{p_d}{4},\frac{p_d}{4},\frac{p_d}{4}\}$.
Here, we set $p_b=p_d=0.2$ (see~\cite{supp} for results when varying $p_b$ and $p_d$).

In the second experiment, we consider $\cC_{\rho^*}$ as the early-stopping process for VQE, taking into account the output of the VQE circuit after 10 and 100 SPSA iterations for $\text{H}_2$ and LiH, respectively.
The noisy VQE data are generated at different SPSA random seeds.
For both experiments, we construct the QAE[2,1,2] for $\text{H}_2$ and QAE[4,1,4] for LiH with $L=1$ ansatz block.
At each bond length, we train the QAE with 200 pairs of noisy VQE data.
%For each bond length, we train the QAE with 200 pairs of noisy VQE data generated at different SPSA optimizer random seeds.
We generate 1000 noisy samples and test with 200 samples of the highest energy.

%static_noise_sel_200.pdf
\begin{figure}
		\includegraphics[width=8.7cm]{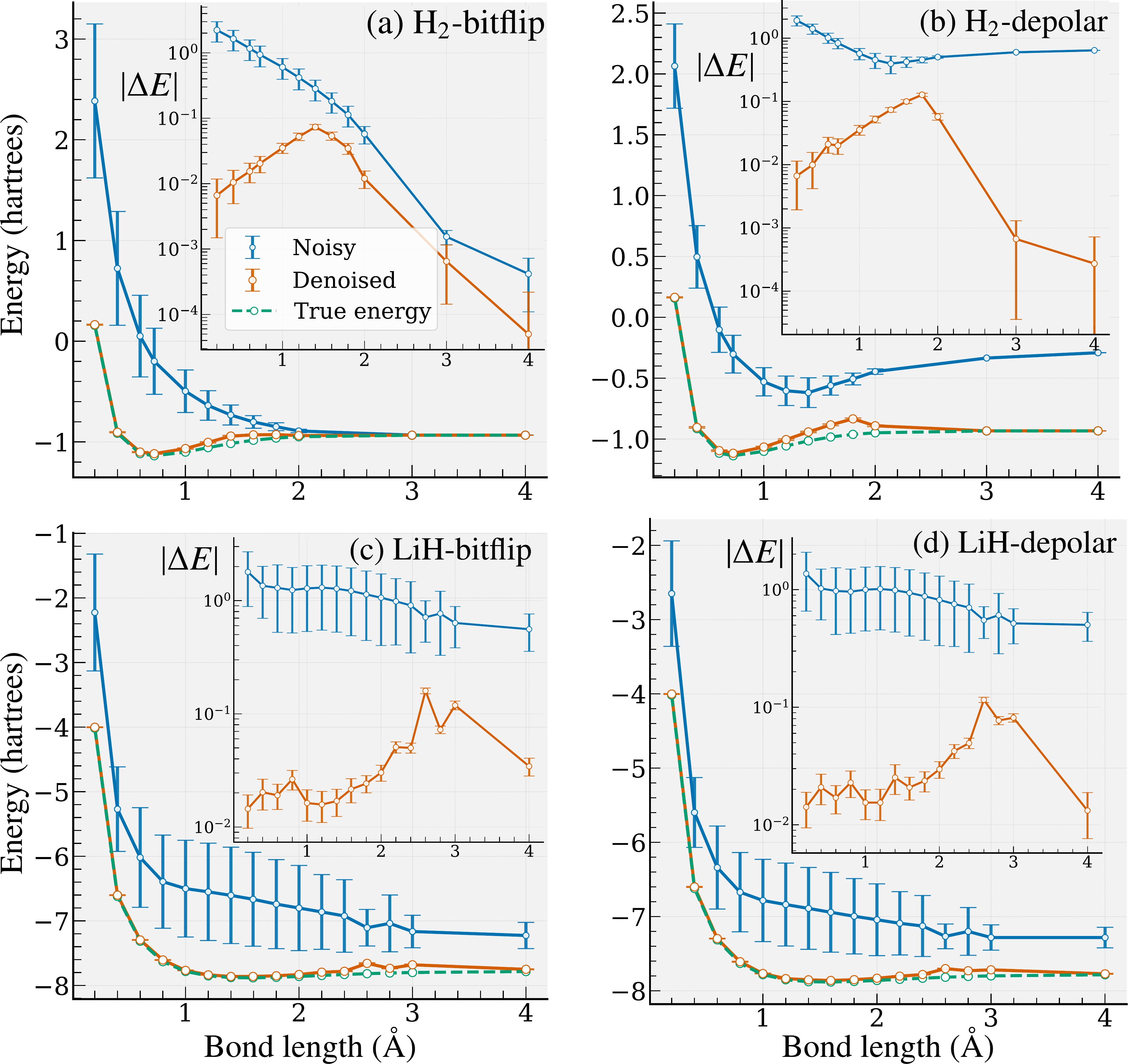}
		\protect\caption{The energy of noisy VQE data and denoised data for $\text{H}_2$ and LiH molecules at each bond length and each type of noise (bitflip, depolar). In the bitflip channel, each qubit was flipped with a probability of 0.2. In the depolarization channel, a single-qubit depolarizing operation was applied to each qubit with a probability of 0.2.
Inset figures describe the energy difference $|\Delta E|$ of these energies with the  ground state energy. Solid lines and error bars describe the average value and standard deviation over 200 test samples.
		\label{fig:res:staticnoise}}
\end{figure}

%H2_LiH_opt_noise_sel_200.pdf
\begin{figure}
		\includegraphics[width=8.7cm]{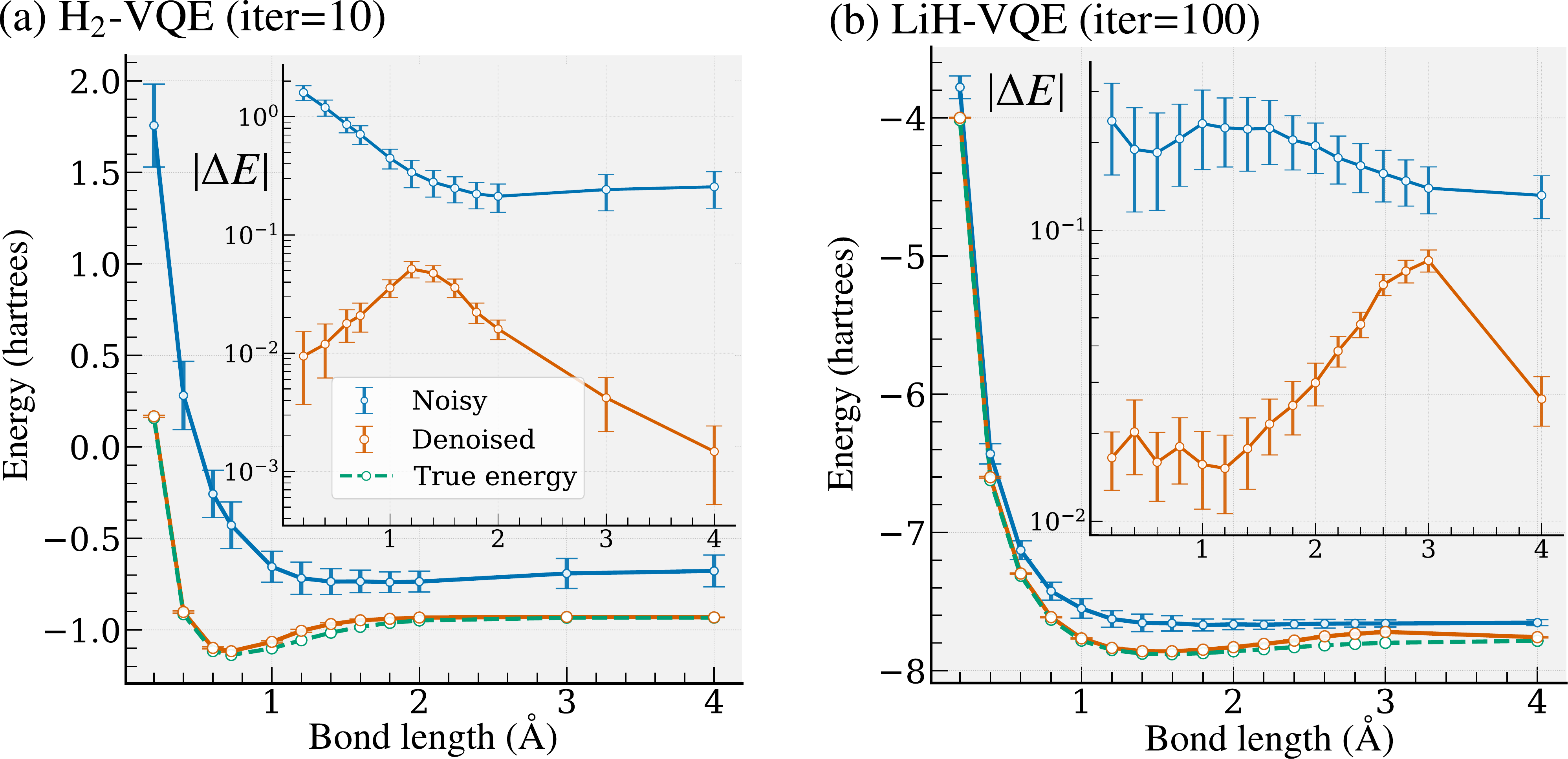}
		\protect\caption{The energy of the noisy data at early steps of VQE optimization and denoised data for (a) $\text{H}_2$ (10 iterations) and (b) LiH molecules (100 iterations).
Inset figures describe the energy difference $|\Delta E|$ of these energies with the  ground state energy.
Solid lines and error bars describe the average value and standard deviation over 200 test samples.
		\label{fig:res:optnoise}}
\end{figure}

Figures~\ref{fig:res:staticnoise} and \ref{fig:res:optnoise} show the energy of noisy and denoised data at various bond lengths.
The inset figures depict the energy errors $|\Delta E|$ of the original and denoised energies with the ground state energy.
Interestingly, even when using noisy data for training, the average performance of our method improves the ground state energy estimation with small fluctuation. 
%Even with noisy data from the early termination of the VQE routine, QAE can reduce energy error and the deviation in energy estimation while increasing the fidelity to the ground state.
%The variance in the estimation can be further reduced by using the parametric controlled-rotation ansatz in the entangling gates~\cite{supp}.
%The decreased fidelity for bond lengths longer than 2$\mathring{A}$ is attributed to the small energy gap between the ground state and the first excited state, and it is difficult to isolate the ground state representation.

Next, we demonstrate that QAE can be used for non-trivial states with more complexity. We apply QAE to the VQE data of one-dimensional (1D) Transverse Field Ising Model (TFIM) of $N$ spins.
The Hamiltonian is defined as
$
H = - \sum_{i} J_{i}\sigma^z_i \sigma^z_{i+1} - \sum_i g_i\sigma^x_i,    
$
where $\sigma^x_i$ and $\sigma^z_i$ are Pauli X and Z matrices acting on the $j$-th spin, $J_i$ is the exchange coupling, and $g_i$ is the transverse magnetic field along the $x$ axis.
We consider uniform interaction $J_{i}=1$ and uniform onsite potential $g_i=g > 0$ with open boundary condition.
This model admits two phases: ordered ($g < 1.0$) and disordered phases ($g > 1.0$), depending on whether the ground state breaks or preserves the $\prod_j\sigma^x_j$ spin-flip symmetry. 
%When $g=1.0$, the system undergoes a quantum phase transition.
The noisy data are generated from the VQE circuit at early stopping optimization with $4N$ iterations of SPSA.
We construct the QAE[$N,1,N$] with $L=3$ ansatz blocks.
We train the QAE with 100 pairs of noisy samples and test with 1000 noisy samples.

\begin{figure}
		\includegraphics[width=8.7cm]{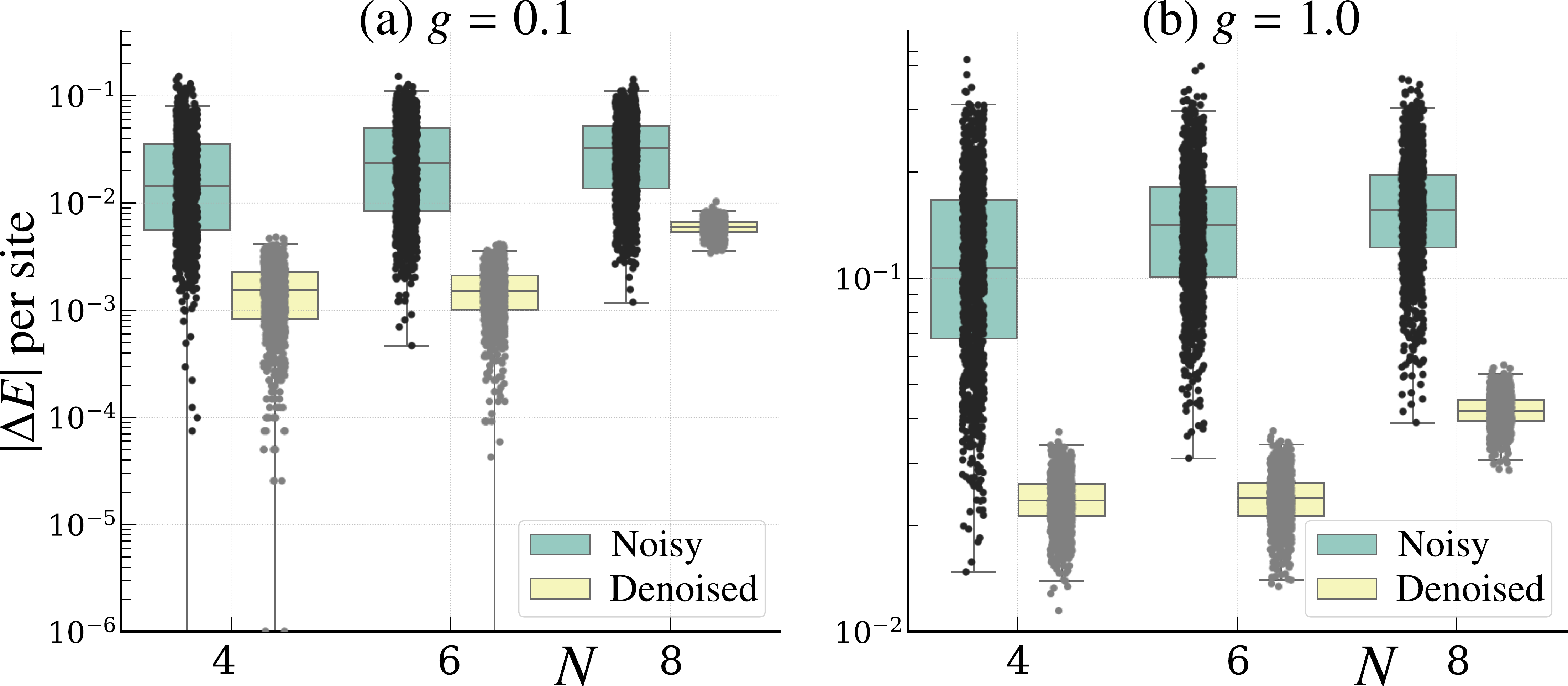}
		\protect\caption{Box and jitter plots displaying the distribution of the energy error $|\Delta E|$ over 1000 test data of 1D TFIM at $N=4,6,8$ spins. The interquartile range is contained within the box, and the 5th and 95th percentiles are marked by the whiskers. The median is the line across the box.
		\label{fig:Ising:optnoise}}
\end{figure}

Figure~\ref{fig:Ising:optnoise} demonstrates the
distribution of the energy error $|\Delta E|$  over 1000 test data of 1D TFIM  at $N=4,6,8$ spins and $g=0.1, 1.0$ (when the ground state has a large entanglement entropy). 
While increasing $N$ can make the problem harder, our method can reduce the error by nearly an order of magnitude~\cite{supp}.

%Ising-Line_sub_1000_TL_RY_CZ_C_3_seed_0_shots_1000_train_100_0_100

%SU2_LAYER_TL_RY_CZ_C_1_layers_4_1_4.pdf
\begin{figure}
		\includegraphics[width=8.7cm]{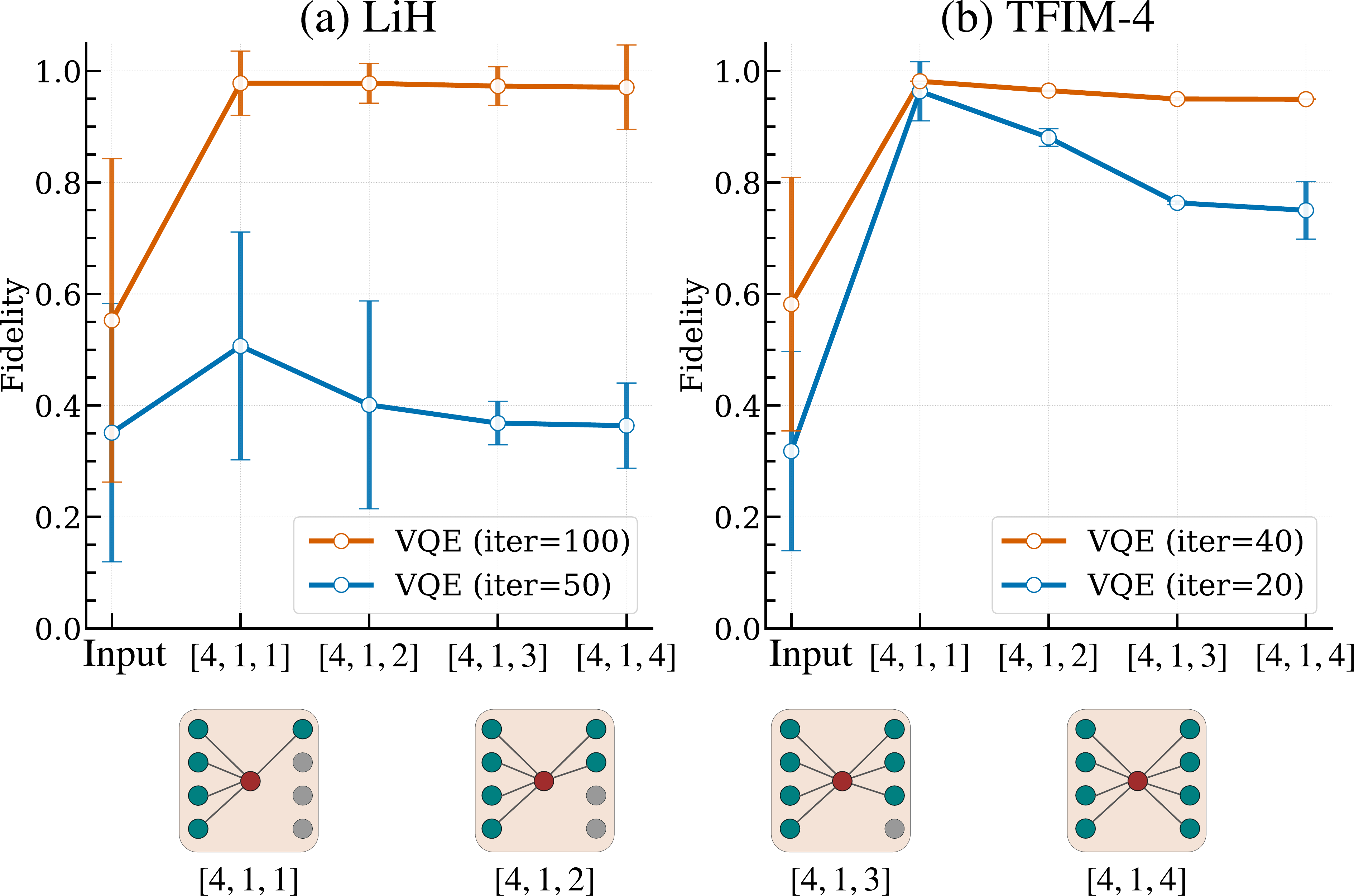}
		\protect\caption{Fidelities between the input and outputs of subsystems of QAE[4,1,4] and the corresponding parts in the ground state of (a) LiH and (b) TFIM-4. The noisy data are generated from VQE circuits at different SPSA iterations. The gray node represents inactive neurons, where subsystems [4,1,1], [4,1,2], and [4,1,3] correspond the subsystems with the first qubit, the first two qubits, the first three qubits of the ground state, respectively. 
  %As the noise level increases, the noisy parts leak out from the bottleneck layer, preventing the decoder from reconstructing the ground state.
   Solid lines and error bars describe the average value and standard deviation over 1000 test samples.
		\label{fig:layers}}
\end{figure}

Finally, to comprehend the mechanism of denoising QAE, we investigate the fidelities between the outputs of subsystems of the QAE and the corresponding parts in the target state as data is forwarded. 
We consider noisy VQE data at 50 (high noise) and 100 (low noise) SPSA iterations of LiH at a bond length of 1.4$\mathring{A}$, and at 20 (high noise) and 40 (low noise) SPSA iterations of TFIM for $N=4$ (TFIM-4) at $g=1.0$.
The QAE[4,1,4] with $L=1$ ansatz block is trained with the same condition in the previous experiments and tested on 1000 noisy data samples.
As shown in Fig.~\ref{fig:layers}, when the input state is forwarded through the QAE, a portion of the target state is reconstructed in the subsystem's output.
The VQE output states at subsystems [4,1,1], [4,1,2], [4,1,3], and [4,1,4] correspond to the reconstructed first qubit, two first qubits, three first qubits, and finally all qubits of the target state.
In a successful training with low noise-level data (orange lines), the dissipative property of the QAE allows it to retain partial information of the target state at the bottleneck neuron and sequentially reconstruct other parts of the target state.
However, if we train the QAE with high noise-level data where the VQE output states have a low overlap with the ground state, irrelevant information leaks out at the bottleneck, preventing the decoder from sequentially reconstructing the ground state (blue lines).

\textit{Conclusion.---}We proposed a method for learning and separating noise from VQE noisy data. It learns to discard noisy and irrelevant information while compressing relevant information into a reduced space to improve the overlap between the output of the VQE and the true target state.
Since the training of QAE is independent of the Hamiltonian in the  VQE problem, it can complement VQE by refining the intermediate state during the optimization.
Our method is especially suited to NISQ devices, where noisy quantum data is common and access to the noiseless data is restricted. It can pave the way for further research into the advantages of quantum models in learning quantum datasets~\cite{schatzki:2021:QMLdata,nakayama:2023:VQEdata}.
Another intriguing aspect is the potential use of our method to connect hybrid quantum systems. When different output states from various algorithms for the same problem are obtained from different platforms, our method can produce a stable result on a robust platform.

\begin{acknowledgments}
The authors acknowledge Koki Chinzei, Yuichi Kamata, Mikio Morita, and Yasuhiro Endo for fruitful discussions.
\end{acknowledgments}

%\bibliography{main.bib}
\providecommand{\noopsort}[1]{}\providecommand{\singleletter}[1]{#1}%

\end{document}

\title{Supplementary Material for\\
``Variational Denoising for Variational Quantum Eigensolver''}

\author{Quoc Hoan Tran}
\email{tran.quochoan@fujitsu.com}
\affiliation{Quantum Laboratory, Fujitsu Research, Fujitsu Limited, Kawasaki, Kanagawa 211-8588, Japan}

\author{Shinji Kikuchi}
%\email{skikuchi@fujitsu.com}
\affiliation{Quantum Laboratory, Fujitsu Research, Fujitsu Limited, Kawasaki, Kanagawa 211-8588, Japan}

\author{Hirotaka Oshima}
\affiliation{Quantum Laboratory, Fujitsu Research, Fujitsu Limited, Kawasaki, Kanagawa 211-8588, Japan}

\date{\today}

\begin{abstract}
This supplementary material describes in detail the calculations, the experiments presented in the main text, and the additional figures.
The equation, figure, and table numbers in this section are
prefixed with S (e.g., Eq.\textbf{~}(S1) or Fig.~S1, Table~S1),
while numbers without the prefix (e.g., Eq.~(1) or Fig.~1, Table~1) refer to items in the main text.
\end{abstract}

\maketitle
\tableofcontents

\section{Training Quantum Autoencoder}
This section explains the quantum circuit-based implementation of Quantum Autoencoder (QAE) and the entire circuit for training. The concept of QAE employed in this study is referenced from Ref.~\cite{bondarenko:2020:prl:QAE} with the circuit implementation based on Qiskit library~\cite{Qiskit} in Ref.~\cite{apache:2020:QAE}.
However, the previous implementation does not provide the hardware-efficient ansatz, which can be run on real NISQ computers. In this implementation, the number of measurements for training scales with the total number of parameters. Furthermore, the number of parameters exponentially increases with the number of qubits used in the system.
In our implementation, we provide a hardware-efficient ansatz circuit for QAE, which can significantly reduce the number of parameters for training. We further use the simultaneous perturbation stochastic approximation (SPSA)~\cite{spall:1999:SPSA} optimization to train these circuits where the number of measurements is independent of the total number of training parameters.

\subsection{Circuit Design}\label{sec:1A:design}
\begin{figure}
		\includegraphics[width=17cm]{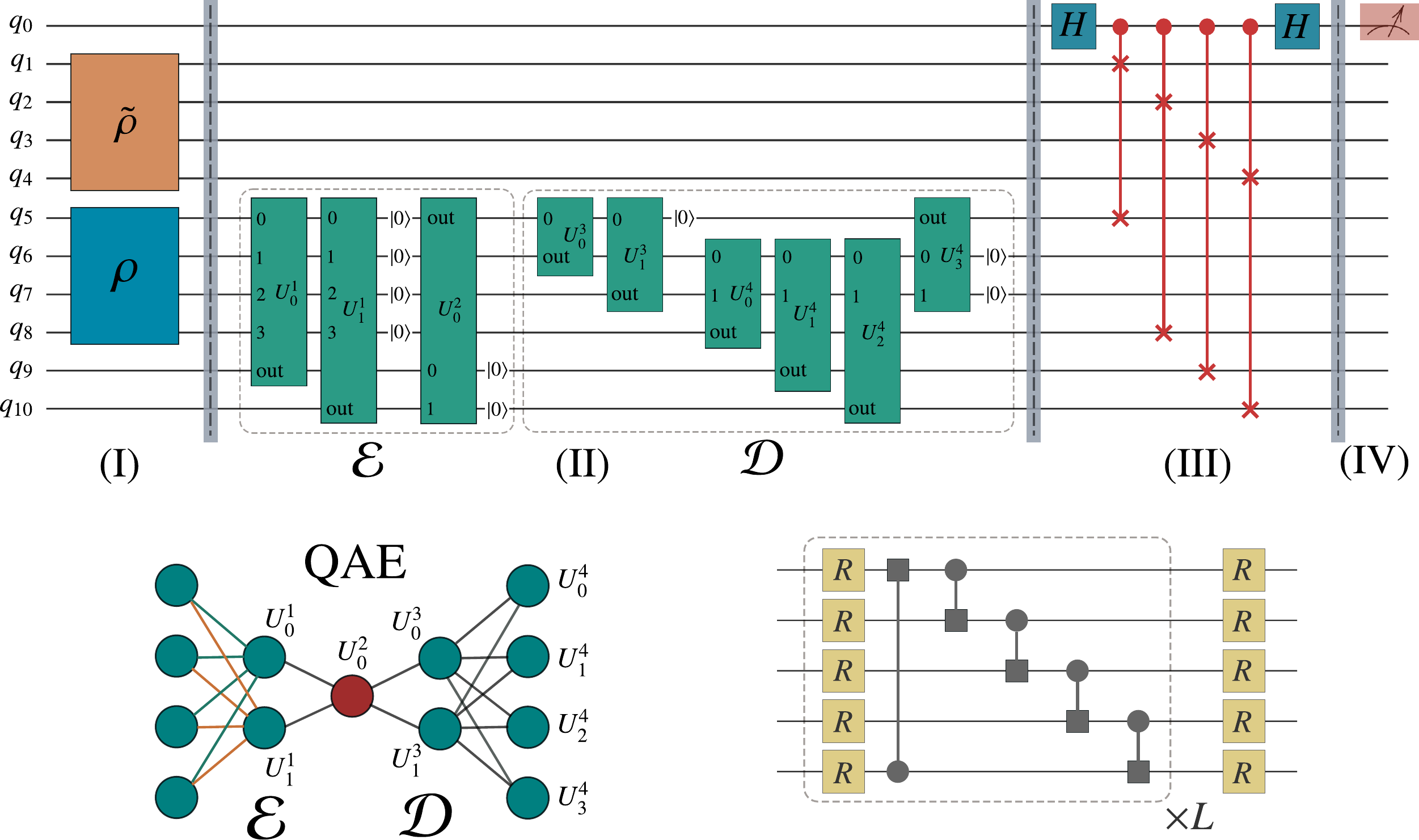}
		\protect\caption{Circuit implementation for training QAE. The quantum circuit used to train the QAE using noisy training data is described in the upper panel. We adopt the concept of implementation in Refs.~\cite{bondarenko:2020:prl:QAE,apache:2020:QAE}. The circuit composes of four parts separated with vertical grey lines: (I) circuit to prepare the input state $\rho$ and the reference state $\tilde{\rho}$ (note that these states are noisy states obtained from other VQE circuits); (II) circuit to implement the abstract structure of QAE with encoder $\cE$ and decoder $\cD$ (here, this circuit represents for the left bottom structure of QAE with $[4, 2, 1, 2, 4]$ nodes in layers); (III) circuit with two Hadamard gates and four SWAP gates to calculate the fidelity between the output of QAE and the target state $\tilde{\rho}$; and (IV) the measurement part. In QAE, each unitary block $U_j^i$ is implemented by a parameterized ansatz circuit consisting of $L$ blocks and final parametric rotation gates $R$ on every single qubit as illustrated in the right bottom figure. Here, each block includes parametric rotation gates on every single qubit and two-qubit entangling gates. In our experiments, we employ the rotation operator of Pauli Y gate as $R_Y(\theta) = e^{-iY\theta/2}$.
  The entangling gates can be nonparametric controlleged gates (controlled-X or controlled-Z) or parametric controlled-rotation gates (controlled-RX or controlled-RZ), which are circularly arranged with the indices of the qubits. They have controlled qubits in gray circles and operate unitaries in gray squares.\label{fig:qae:circuit}}
\end{figure}

The upper panel of Fig.~\ref{fig:qae:circuit} describes the circuit design used to train the QAE.
Since the QAE in our study takes an input state $\rho$ as a noisy state and trains the parameterized circuits to make the output state as close as possible to another noisy state $\tilde{\rho}$, the whole circuit for training consists of four parts: (I) circuit for state preparation for both input state $\rho$ and target state $\tilde{\rho}$ (in our experiments, the input and reference state are pure states prepared by VQE circuits), (II) circuit of QAE, (III) circuit to compute the fidelity of the output of QAE and the reference state $\tilde{\rho}$, and (IV) measurement part. The fidelity of the density matrix output of QAE and the reference pure state is calculated as $2p_0 - 1$, where $p_0$ is the probability of obtaining 0 in  the measurement.

Figure~\ref{fig:qae:circuit} shows the circuit implementation for the QAE structure with $[4, 2, 1, 2, 4]$ nodes in layers to denoise $4$-qubit states (right bottom figure).
Each neuron corresponds to a qubit, and unitary circuits connect neurons in subsequent coupled layers.
Qubits in each layer are dissipative after the forward process in the next layer, making it possible to construct the reduced space for QAE.
Consider the QAE with structure $[m_1, \ldots, m_{M}]$ ($M\geq 2$).
For neuron $j$th in layer $i+1$, we denote $U_j^i$ as the parameterized unitary acting on this neuron and the neurons on the preceding layer.
These unitary matrices are subject to optimization.
The unitary between layer $i$ and $i+1$ can be expressed as $U^i=\prod_j U_j^i$.
We denote $\beta_i$ as the state at $i$th layer with $m_i$ neurons. Consequently, the transformation between $\beta_{i}$ and $\beta_{i+1}$ can be expressed through the quantum channel
\begin{align}
    \beta_{i+1} = \cQ_i(\beta_{i}) = \tr_i\left[ U^i(\beta_i \otimes (\ket{0}\bra{0})^{\otimes m_{i+1}})(U^{i})^{\dagger} \right]. 
\end{align}
Here, $\cQ_i$ extends an $m_i$-qubit state into an $(m_i+m_{i+1})$-qubit state, then reduces it into an $m_{i+1}$-qubit state via tracing out qubits in layer $i$.
In this way, with topology [4,2,1,2,4], the encoder $\cE=\cQ_1\cQ_2$ comprises three unitaries in two channels, where the first channel $\cQ_1$ extends $4$-qubit states (layer 1) into a $6$-qubit state then reduces it into the $2$-qubit state (layer 2), the second channel $\cQ_2$ further extends $2$-qubit states into the $3$-qubit state then reduces it into the $1$-qubit state (layer 3).
The decoder $\cD=\cQ_3\cQ_4$ does inversely on this $1$-qubit state to reconstruct the final $4$-qubit state with six unitaries.
Consequently, this dissipative design of QAE requires $\max\{(m_i+m_{i+1})\}$ qubits and the training circuit requires $1 + m_1 + \max\{(m_i+m_{i+1})\}$ qubits with $m_1 + 1$ qubits used to calculate the fidelity.

The authors in Refs.~\cite{bondarenko:2020:prl:QAE,apache:2020:QAE} consider the optimization on all possible values of each unitary $U_j^i$ (applied to the $j$th neuron on layer $i+1$ and  the neurons on the preceding layer) in circuit part (II). Here, $U_j^i$ is represented as  
\begin{align}\label{eqn:full:uni}
    U_j^i = e^{iK_j^{i}} \text{ where }  K_j^{i} = \sum_{\sigma \in\{I, X, Y, Z\}^{\otimes (m_i + 1)}} k_{\sigma}\sigma,
\end{align}
where the Hermitian matrix $K_j^{i}$ is uniquely defined by $4^{m_i + 1}$ coefficients $k_\sigma$ on the expansion in the Pauli basis $\sigma \in\{I, X, Y, Z\}^{\otimes (m_i + 1)}$ for the real vector space of $2^{m_i+1} \times 2^{m_i+1}$.
Thus, for the QAE with structure $[m_1, m_2, \ldots, m_{M}]$, the total number of training parameters is
\begin{align}
    P = \sum_{i=1}^{M-1} m_{i+1}4^{m_i+1},
\end{align}
which is exponentially scaling with the number of qubits used in the system.  Consequently, we need to explore a large number of training parameters, which makes the training phase time-consuming and challenging.
Furthermore, in principle, the implementation of the representation in Eq.~\eqref{eqn:full:uni} in real NISQ devices is inefficient.

In our research, we provide an ansatz-based implementation for QAE, where each unitary block $U_j^i$ is implemented by a parameterized quantum circuit comprising $L$ blocks and final parametric rotation gates $R$ on every single qubit (the right bottom of Fig.~\ref{fig:qae:circuit}). Here, each block includes parametric rotation gates on every single qubit and two-qubit entangling gates. In our experiments, we use the rotation operator of the Pauli Y gate as $R_Y(\theta) = e^{-iY\theta/2}$. The non-parametric two-qubit entangling gates are controlled-X or controlled-Z (controlled qubits in gray circles and operate unitaries in gray squares), which are arranged circularly with indexes of qubits.
Our implementation is hardware efficient and can reduce the number of parameters for training (see Table~\ref{tab:params}).
The number of parameters only scales linearity with the number of qubits and layers used in each ansatz block, given the same depth of QAE.  

\begin{table*}
	\caption{\label{tab:params} The number of parameters used in the earlier implementation in Refs.~\cite{bondarenko:2020:prl:QAE,apache:2020:QAE} and our implementation for QAE. Here $L$ is the number of ansatz blocks used in each unitary of QAE (see the right bottom figure in Fig.~\ref{fig:qae:circuit}).}
	\begin{ruledtabular}
    \begin{tabular}{l c c c c} 
	Structure & Implementation in Ref.~\cite{apache:2020:QAE} & Ours (L=1) & Ours (L=2) & Ours (L=3) \\ \hline
 \textrm{[2, 1, 2]}     & 96 & 14  & 21 & 28  \\
 \textrm{[3, 1, 3]}     & 304 & 20  & 30 & 40 \\
 \textrm{[4, 1, 4]}     & 1088 & 26  & 39 & 52 \\
 \textrm{[4, 2, 1, 2, 4]}     & 2400 & 58 & 87 & 116 \\
    \end{tabular}
	\end{ruledtabular}
\end{table*}

\subsection{Training Algorithm}
The training in denoising QAE can be formulated to maximize the empirical cost
\begin{align}\label{eqn:emp:denoise}
    \cL(\bth) = \dfrac{1}{N}\sum_{i=1}^N\cF(\rho^A_i, \cD(\cE(\rho^B_i)),
\end{align}
given $N$ paired training states $(\rho^A_i, \rho^B_i)$, 
where $\bth=(\theta_1,\ldots,\theta_P)$ is the parameter vector with $P$ parameters used to construct $\cE$ and $\cD$.
The paired training states $(\rho^A_i, \rho^B_i)$ are sampled from a noisy process $\cC_{\rho^*}$ where we do not know the true target state $\rho^*$. Therefore, this is different from the training of classical autoencoder, where only pairs of identical data are considered for training.

A typical method to maximize the cost in Eq.~\eqref{eqn:emp:denoise} is the vanilla gradient descent (GD) method, which has been employed to train QAE in Ref.~\cite{apache:2020:QAE}.
Here, parameters are updated over iterations, where the update is based on the gradient of the cost function to increase the cost. At $(t+1)$-th iteration step, GD updates the parameter $\theta_p$  as 
\begin{align}\label{eqn:update:params}
    \theta^{(t+1)}_p = \theta^{(t)}_p +  \eta \dfrac{\partial \cL}{\partial \theta_p}(\bth^{(t)}),
\end{align}
where $\eta$ is the learning rate.
At each iteration, the gradient $\dfrac{\partial \cL}{\partial \theta_p}(\bth)$ is evaluated for every index $p$. Considering the sufficient small perturbation step $\epsilon$, we can use the finite difference method to approximate the gradient as follows:
\begin{align}\label{eqn:grad:finite}
    \dfrac{\partial \cL}{\partial \theta_p}(\bth) \approx \dfrac{\cL(\bth + \bep_p) - \cL(\bth)}{\epsilon}, 
\end{align}
where $\bep_p$ is an indicator vector with $\epsilon$ at the $p$-th position in the parameter vector $\bth$, and 0 otherwise.
From Eq.~\eqref{eqn:update:params} and Eq.~\eqref{eqn:grad:finite}, we can estimate the complexity in terms of the number of measurements for each iteration as $N(P+1)\times S$, where $N$ is the number of training data, $P$ is the number of training parameters, and $S$ is the number of shots to run the circuit to estimate the probability of obtaining $0$ in the measurement part to compute the fidelity.
Since $N$ and $P$ are often fixed, the number of quantum measurements heavily depends on the number of parameters in the model, which is what caused the major constraint in the prior approach in Ref.~\cite{apache:2020:QAE}.

To overcome this limitation, we employ the SPSA~\cite{spall:1999:SPSA} optimizer to replace the gradient estimation dimension by dimension with a stochastic one that needs only two evaluations of the cost function regardless of the number of parameters.
SPSA does not provide a stepwise unbiased gradient estimation but is appropriately effective for large-scale optimization problems in the presence of noise. Thus, it is recommended in the optimization routine with NISQ devices or noisy quantum simulators.

In SPSA, a $P$-dimensional random direction $\bb$ is sampled from a $P$-dimensional discrete uniform distribution on $\{-1, 1\}^P$. Along this sampled direction, the gradient component is approximated with a finite difference method with two evaluations of the cost function as
\begin{align}
    \boldsymbol{\nabla}_{\bb}\cL(\bth)  \equiv \bb \cdot \boldsymbol{\nabla}\cL(\bth) \approx \dfrac{\cL(\bth + \epsilon\bb) - \cL(\bth - \epsilon\bb)}{2\epsilon}.
\end{align}
Therefore, the stochastic estimator of the gradient is constructed as $g_{\bb}(\bth)=\left( \boldsymbol{\nabla}_{\bb}\cL(\bth) \right) \bb$, which gives the update rule:
\begin{align}\label{eqn:SPSA:update}
    \bth^{(t+1)} = \bth^{(t)} + \eta g_{\bb^{(t)}}(\bth^{(t)}),
\end{align}
where $\bb^{(t)}$ is sampled at $t$-th iteration.
Furthermore, to reduce the fluctuation in the stochastic estimator $g_{\bb}(\bth)$, we consider multiple samples of $\bb$ at each iteration and then take the average value for this estimator.
In our experiments, we use two different samples of $\bb$ at each iteration.
Therefore, the number of measurements for each iteration is $4NS$, where we use $N=200$ training data and the number of shots $S=1000$ in all simulations.
 
To reduce the fluctuation and enhance the convergence of the optimization process, we have incorporated the concept of AMSGrad optimizer~\cite{reddi:2018:AmsGrad} into the SPSA algorithm. The AMSGrad optimizer considers the maximum of past squared gradients when updating parameters. The transformation of Eq.~\eqref{eqn:SPSA:update} results in the following equations:
\begin{align}
    m_t &= \beta_1 m_{t-1} + (1-\beta_t)g_t,\\
    v_t &= \beta_2 v_{t-1} + (1-\beta_2)g^2_t,\\
    \hat{v}_t &= \max(\hat{v}_{t-1}, v_t),\\
     \bth^{(t+1)} &= \bth^{(t)} + \dfrac{\eta}{\sqrt{\hat{v}_t}+\nu}m_t,
\end{align}
where $g_t,g_{t-1},\ldots$ represent the sequence of past stochastic estimators $g_{\bb}(\bth)$, and $\nu$ is a small non-zero perturbation introduced to prevent division by zero. Here, we define this optimizer as AMSGradSPSA.

In our implementation, we further introduce the mini-batch learning and the schedule for learning rate $\mu$ for more efficient learning. 
Here, mini-batch learning involves training a model on small batches of data rather than using the entire data set at once. In mini-batch learning, the data are divided into smaller batches, and the model is trained on each batch in turn.
Because mini-batch learning exposes the model to a greater variety of data during training,
it can increase the model's generalization.
In our training process, we randomly split the data set into batches of 50 instances at each epoch. We calculate the cost function and gradient at each batch to update the parameters. Consequently, each training epoch requires updating parameters four times. We run all experiments with 100 epochs of training.
Furthermore, the small perturbation $\epsilon$ is set to $\epsilon=0.1$ for gradient estimation.
The learning step at the first epoch is set to $\eta = \eta_0 = 0.1$ but is reduced at the scale of 0.8 every ten training epochs.
Other parameters for AMSGradSPSA are $\beta_1=0.9, \beta_2=0.999$, and $\nu = 10^{-8}$.

\pagebreak
\section{VQE implementation}
In our experiments, we create training data for QAE from VQE in the electronic structure problem in quantum chemistry.
First, the molecular Hamiltonian is mapped to qubit Hamiltonian with via the parity transformation from the second-quantized fermionic Hamiltonian. We can employ reduction techniques such as Z2 symmetries and two-qubit reduction to decrease the number of qubits required.
In our experiments, we simulate $\text{H}_2$, LiH, and $\text{BeH}_2$ molecules using the STO-3G basis with the parity mapping.
The H atom has its electron in the 1s atomic orbitals, while the Li atom has 3 electrons and Be atom has 4 electrons in 1s, 2s and 2p (with $x, y, z$ components) atomic orbitals. When forming the molecules, these atomic orbitals contribute to the formation of molecular orbitals. For our molecular systems, this results in a total of 2, 6 (=5+1), and 7 (=5+2) molecular orbitals for $\text{H}_2$, LiH, and $\text{BeH}_2$, respectively.
In terms of spin orbitals (accounting for spin-up and spin-down), this corresponds to 4, 12, and 14 spin orbitals in the Hamiltonian for molecular $\text{H}_2$, LiH, and $\text{BeH}_2$, respectively. Then, the original Hamiltonians for $\text{H}_2$, LiH, and $\text{BeH}_2$ would require 4, 12, and 14 qubits to represent, respectively. In the subsequent steps, we follow Ref.~\cite{kandala:2017:nature:VQE} and the explanation in IBM Quantum Challenge 2021~\cite{ibm:challenge:2021} to reduce these into a more compact problem without significantly affecting the target ground state energy.

For $\text{H}_2$, the parity mapping removes the two qubits associated with the spin parities of the system, subsequently encoding the Hamiltonian into a two-qubit system. For BeH2, we select an active space comprising four molecular orbitals. We then map the Hamiltonian of the eight spin-orbitals using parity mapping. This process eliminates two qubits, reducing the computation to a six-qubit problem.

For LiH, the original problem requires 12 qubits. We begin the reduction by freezing the core electrons in the Li 1s orbital. As these electrons do not significantly influence the chemistry, this step allows us to eliminate two qubits. In the ground state of LiH, the primary interactions involve the H 1s and Li 2s orbitals. However, there's also some involvement of the $2p_z$ orbital due to the molecule's orientation. Recognizing this, we can exclude the molecular orbitals corresponding to $2p_x$ and $2p_y$. This removal further reduces the system by two qubits. To continue the reduction, we employ the parity mapping combined with a two-qubit reduction, leading to a reduction of another two qubits. This means that the Hamiltonian of LiH can now be represented using just six qubits. By further leveraging the symmetries within the Hamiltonian, and recognizing that it commutes with both spin-up and spin-down number operators, we can remove one qubit for each symmetry, thus eliminating two more qubits. After all these steps, the LiH problem is effectively condensed to a four-qubit system.

The primary goal of the VQE circuit is to accurately represent the ground state wavefunction through a series of single-qubit rotations and entangling operations. A distinct advantage of quantum computers is their ability to efficiently represent and store exact wavefunctions. For systems encompassing more than a few atoms, such a representation becomes computationally prohibitive on classical computers. For chemical systems, we have two prominent types of ansatzes to choose from: q-UCC ansatzes~\cite{sokolov:2020:qUCC} and heuristic ansatzes that emphasize hardware efficiency, like the two-local ansatz~\cite{barkoutsos:2018:twolocal}. While q-UCC ansatzes draw inspiration from physical processes, aiming to map electron excitations to quantum circuits, their need for many layers and gates often renders them less efficient. In our research, we focus on the heuristic ansatzes for the VQE problem.
The VQE quantum states are prepared by the following two-local variational ansatz state~\cite{barkoutsos:2018:twolocal} of parameter $\bzet$:
\begin{align}~\label{VQE:circuit}
    \ket{\psi(\bzet)} = \prod_{d=1}^D \left( \prod_{q=1}^Q \left[ U_{R}^{q,d}(\bzet) \right] \times U_{\text{Ent}} \right) \times  \prod_{q=1}^Q\left[U_{R}^{q,0}(\bzet)\right]\ket{00\ldots 0}, 
\end{align}
for $Q$ qubits consisting of $D$ entangling gates $U_{\text{Ent}}$ alternating with $Q(D+1)$ rotation gates on each qubit.
In our case, we begin with the zero state $\ket{0}$ instead of the Hartree-Fock state. For the heuristic ansatz of VQE in our experiments, we use $U_R(\bzet) = R_Y(\bzet_1)R_Z(\bzet_2)$, and  $U_{\text{Ent}}$ is composed of CNOT gates placed in linear with indexes $(k,k+1)$ of qubits. The number of parameters in this circuit is $Q(2D+1)$.
In the main text, we construct the variational circuits with $D=1$ and optimize the circuits using Qiskit's implementation of SPSA.

\section{Denoising for VQE of Molecular Hamiltonians}

This section presents additional results for denoising the VQE data of molecular Hamiltonians.

\subsection{Demonstration of denoising}
\begin{figure}
		\includegraphics[width=17cm]{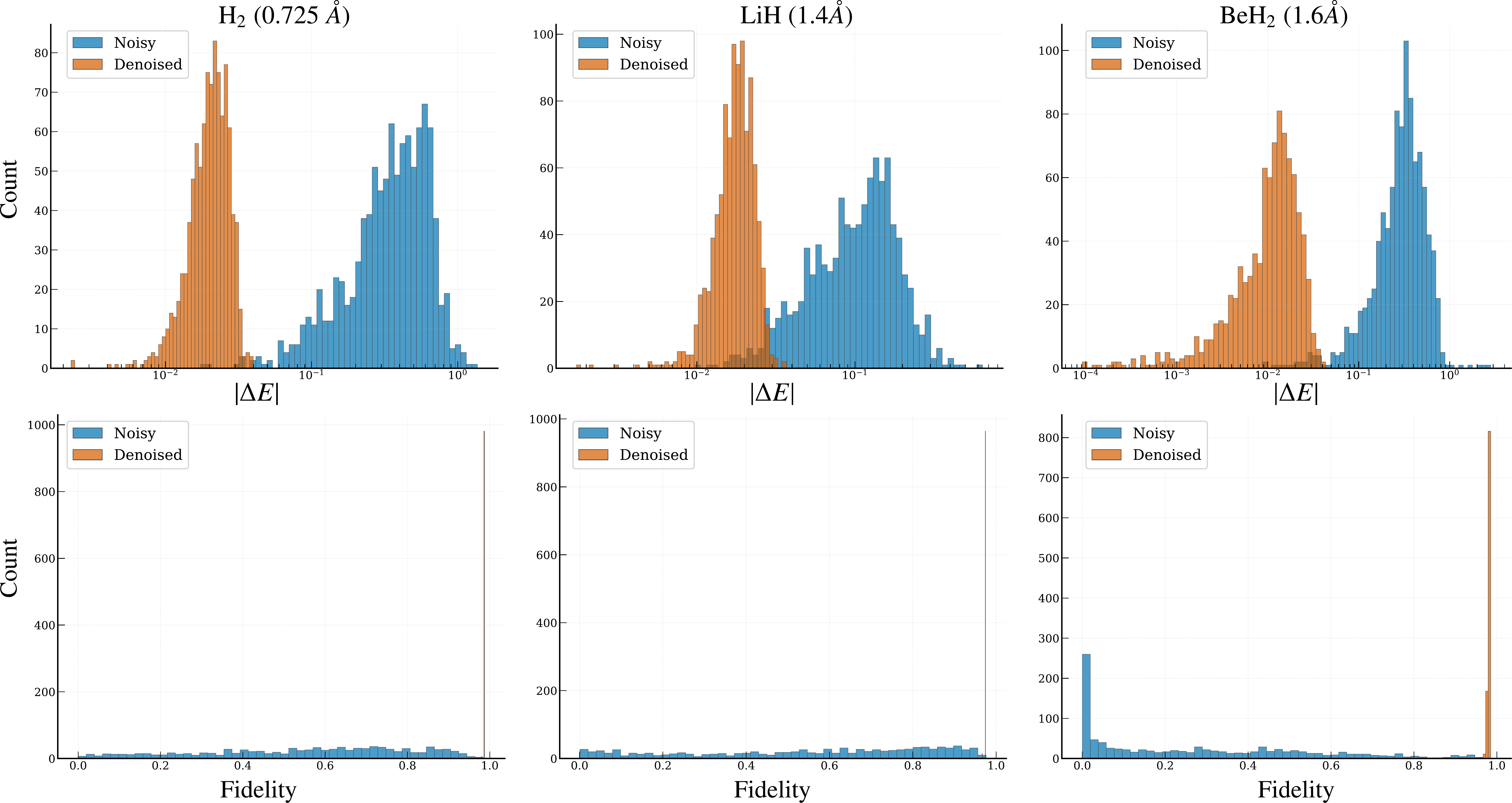}
		\protect\caption{The histogram of energy difference $|\Delta E|$ with the ground state energy and the histogram of the fidelity with the ground state of the noisy and denoised VQE data for $\text{H}_2$, LiH, and $\text{BeH}_2$ molecules at the bond length 0.725$\mathring{A}$, 1.4$\mathring{A}$, and 1.6$\mathring{A}$, respectively.
  QAE is trained with 200 pairs and tested with 1000 noisy VQE samples of the early stopping at 10, 100 and 500 SPSA iterations for $\text{H}_2$, LiH, and $\text{BeH}_2$, respectively.
		\label{fig:opt:test1000}}
\end{figure}

First, we consider the noisy training data as the output of VQE circuits at 10, 100, and 500 SPSA iterations for $\text{H}_2$, LiH, and $\text{BeH}_2$, respectively.
The noisy VQE data are generated at different random seeds of the SPSA optimizer.
We train the QAE with 200 pairs and test with 1000 samples of noisy data.
Here, we employ QAE[2,1,2] for $\text{H}_2$,  QAE[4,1,4] for LiH, and QAE[6,1,6] for $\text{BeH}_2$ with $L=1$ ansatz block.
In QAE, the parametric rotation gate $R$ on each qubit is the rotation of Pauli Y and the entangling gates are nonparametric controlled-Z gates.
Figure~\ref{fig:opt:test1000} depicts the histogram of energy difference $|\Delta E|$ between the estimated energy and the ground state energy, and the histogram of the fidelity with the ground state of the noisy and denoised data for $\text{H}_2$, LiH and $\text{BeH}_2$ molecules, respectively.
The energy estimation and the overlap with the ground state are largely improved in all cases.

\begin{figure}
		\includegraphics[width=12cm]{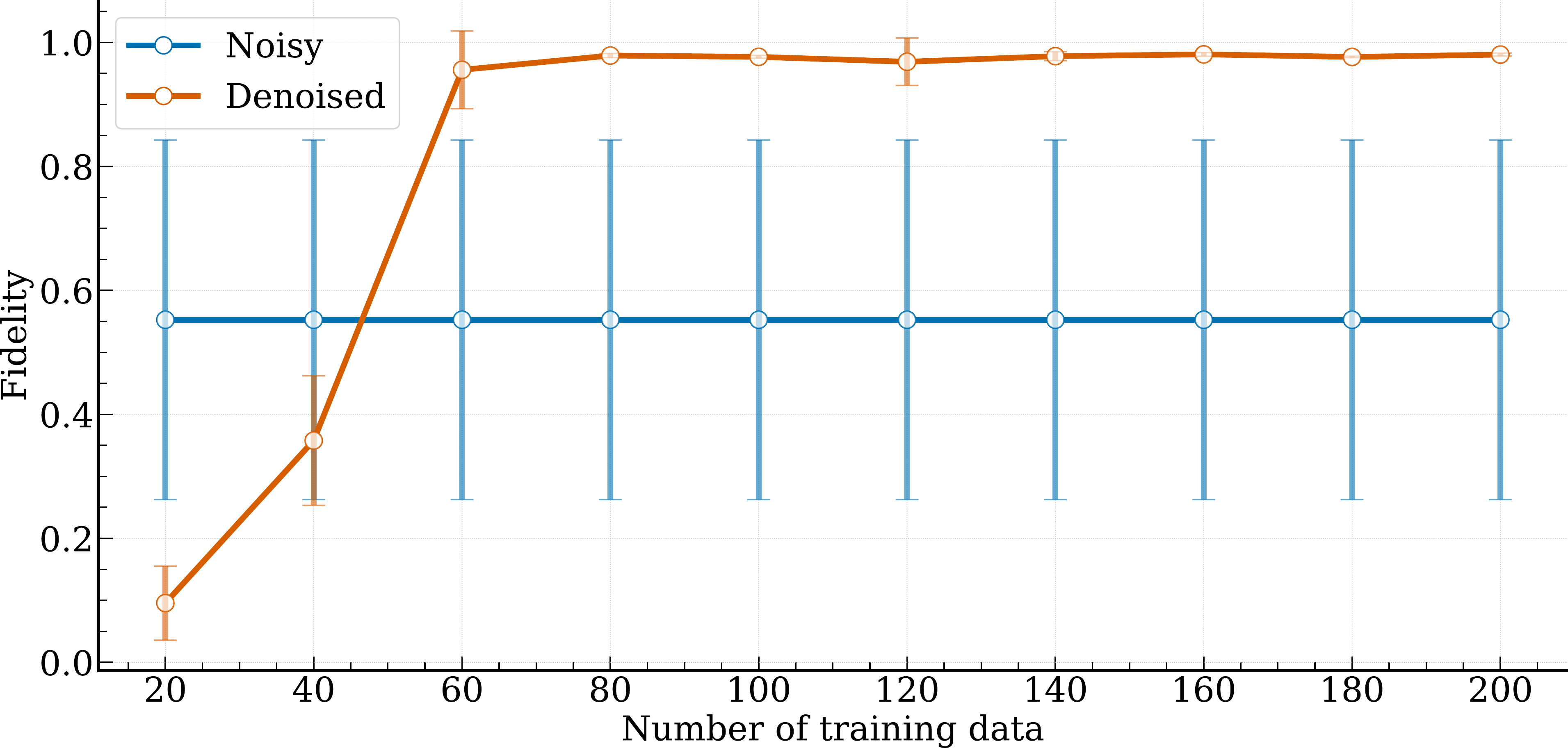}
		\protect\caption{The fidelities of noisy VQE and denoised states with the ground state via the number of training data for LiH molecule ((at a bond length of 1.4$\mathring{A}$). The QAE is trained with 200 pairs and tested with 1000 noisy samples. We consider the VQE states at the early stopping of 100 SPSA iterations as noisy data. We employ QAE[4,1,4] with $L=3$ ansatz blocks, where the parametric rotation gate on each qubit is the Pauli Y rotation, and the entangling gates are nonparametric controlled-Z gates.
		\label{fig:num_train:LiH_BL_1.4}}
\end{figure}

In addition, we analyze the effect of the number of training pairs on the denoising performance of our method for VQE data of LiH. Figure~\ref{fig:num_train:LiH_BL_1.4} shows the fidelity of the denoised states for 1000 test samples as a function of the number of training pairs. We use QAE[4,1,4] for LiH with $L=3$ ansatz blocks. Notably, our method achieves high fidelity even with a small number of training pairs, suggesting the potential for efficient learning.

\begin{figure}
		\includegraphics[width=17cm]{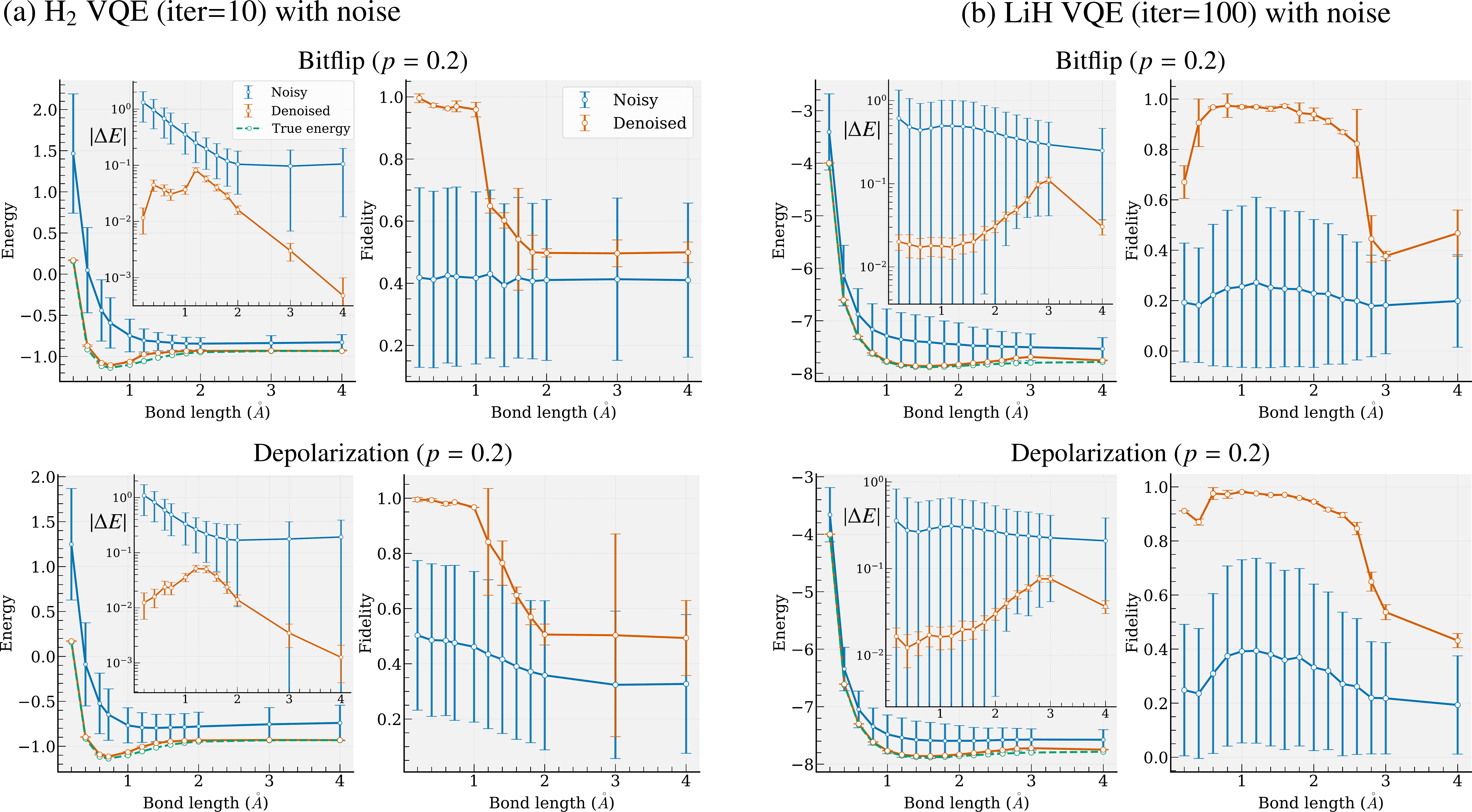}
		\protect\caption{The energy, the energy difference $|\Delta E|$ with the ground state energy(inset figures), and the fidelity with the ground state of the noisy VQE data and denoised data for $\text{H}_2$ and LiH at each bond length.
  Here, noisy VQE data are created with an \textbf{early stopping of SPSA and another noise such as bitflip or depolarization noise}). VQE data at 10 (for $\text{H}_2$) and 100 (for LiH) SPSA iterations for the VQE optimization are subjected to either the bitflip or depolarization channel with a noise strength $p$ of $0.2$. For QAE[2,1,2] (for $\text{H}_2$) and QAE[4,1,4] (for LiH), we use one ansatz block, where the parametric rotation gate $R$ on each qubit is the rotation of Pauli Y gate, and the entangling gates are nonparametric CZ gates.
  Solid lines and error bars describe the average value and standard deviation over 1000 test samples.
		\label{fig:opt-bitflip-depolar}}
\end{figure}

Next, we consider the experiment where we apply bitflip and depolarizing noise channels to the output of VQE circuits at the early stopping of 10 and 100 iterations to generate noisy data for $\text{H}_2$ and LiH.
At each bond length, we train the QAE with 200 pairs of noisy VQE data generated at different random seeds. We further generate 1000 samples of noisy data for testing to see the denoising effect in ground-state estimation. Figure~\ref{fig:opt-bitflip-depolar} shows the energy of the noisy and denoised data and the fidelities with the ground state at each bond length for the $\text{H}_2$ and LiH molecules. Here, we construct the QAE[2,1,2] for $\text{H}_2$ and the QAE[4,1,4] for LiH with $L=1$ ansatz block, where the rotation gate $R$ on each qubit in QAE is the parametric rotation gate of the Pauli Y gate, and the entangling gates are nonparametric CZ gates.

\subsection{Noise robustness in QAE}
In the main text, we did not consider the noise model in the QAE. This is motivated by a strong interest in hybrid systems, where a well-controlled system can be used to remove noise from a noisy quantum platform.

\begin{figure}
\centering
	\includegraphics[width=16cm]{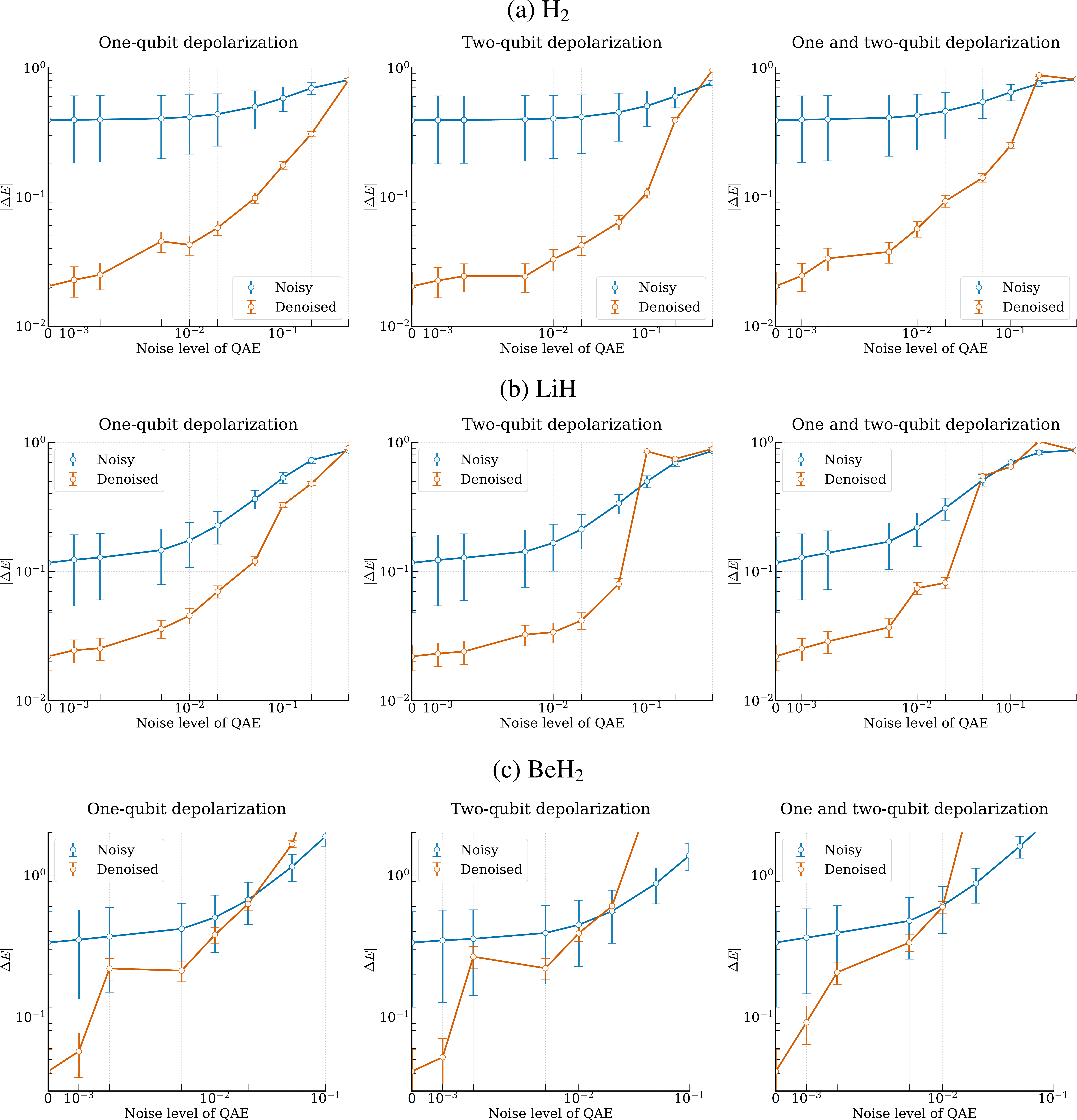}
		\protect\caption{The variation in energy error for the output state obtained using QAE (orange lines) to find the ground state energy of (a) $\text{H}_2$ (at a bond length of 0.725$\mathring{A}$) (b) LiH (at a bond length of 1.595$\mathring{A}$), and (c) $\text{BeH}_2$ (at a bond length of 1.6$\mathring{A}$) as we vary the depolarizing error rate (noise level) in each noise pattern of the QAE's circuit. The circuit used to evaluate the energy of the output state is simulated under the same noise conditions as the QAE, resulting in an increase in energy error in the original noisy data (blue lines).}\label{fig:QAEnoise}
\end{figure}

In this section, we have conducted numerical experiments to investigate the robustness of noise in QAE. We simulated QAE circuits with single-qubit depolarizing errors, two-qubit depolarizing errors, or both. Figure~\ref{fig:QAEnoise} displays the variation in energy error for the output state obtained using QAE (orange lines) to find the ground state energy of $\text{H}_2$ (at a bond length of 0.725$\mathring{A}$), LiH (at a bond length of 1.595$\mathring{A}$), and $\text{BeH}_2$ (at a bond length of 1.6$\mathring{A}$) as we vary the depolarizing error rate (noise level) in each noise pattern of the QAE's circuit. The circuit used to evaluate the energy of the output state is simulated under the same noise conditions as the QAE, resulting in an increase in energy error in the original noisy data (blue lines).
%Similar to the experiments in the main text, for each noise level of QAE, we constructed the QAE[4,1,4] with an $L=1$ ansatz block, where the rotation gate $R$ is the parameterized rotation Pauli Y gate, denoted as $R_Y(\theta) = e^{-iY\theta/2}$, and the non-parameterized two-qubit entangling gates are controlled-Z gates. We trained the QAE with 200 pairs of noisy VQE data and tested it with 1000 samples of noisy VQE data.
Interestingly, VQE can exhibit a denoising effect under a certain tolerance of noise level in the ansatz circuit.

\subsection{Performance of QAE with different noise levels in data}

We investigate whether the QAE can maintain its performance for larger error. In the first experiment, we vary the noise strength in the data with the depolarization noise strength $p_d \in \{0.1, 0.2, 0.3, 0.4, 0.5, 0.6, 0.7, 0.8, 0.9, 1.0\}$ and the bitflip noise strength $p_b \in \{0.1, 0.2, 0.3, 0.4, 0.5\}$. Here, the depolarizing channel is implemented by applying one of the Pauli matrices $\{I, X, Y, Z\}$ to each qubit with respective probabilities $\{1-3p_d/4, p_d/4, p_d/4, p_d/4\}$, and the bitflip channel is implemented by applying $\{I, X\}$ to each qubit with respective probabilities $\{1-p_b, p_b\}$.

\begin{figure}
\centering
		\includegraphics[width=17cm]{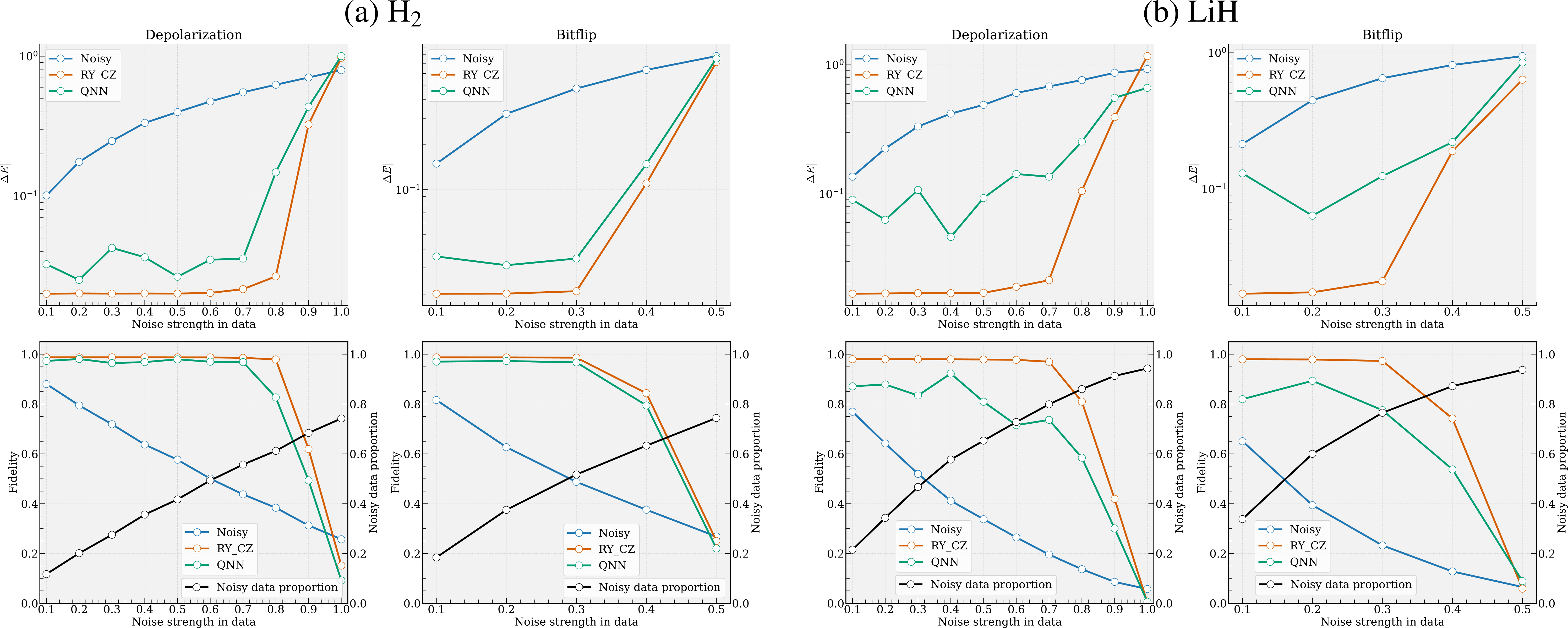}
		\protect\caption{The robustness of the QAE concerning different noise types (depolarization and bitflip) and noise strengths in the final result of VQE for (a) $\text{H}_2$ (at a bond length of 0.725$\mathring{A}$) and (b) LiH (at the bond length 1.4$\mathring{A}$). The solid lines represent the average energy errors or fidelity with the ground state. The QAE is constructed with an ansatz block, where the rotation gates RY are applied to all qubits, and the two-qubit controlled-Z gates are placed circularly with the indices of the qubits (denoted as RY\_CZ). The label QNN represents the implementation of QAE in previous research~\cite{bondarenko:2020:prl:QAE,apache:2020:QAE}.
For each random initialization, we train both structures with 200 random pairs of noisy data for 100 training epochs and test with 1000 samples of the noisy data. The predicted energy error and predicted fidelity are averaged over these 1000 samples and ten times for random initialization. In the fidelity plots, solid black lines represent the average proportion of ``very noisy data" in the training data set. Here, ``very noisy data" is defined as data with fidelity to the ground state of less than 0.2.}~\label{fig:robust:data}
\end{figure}

Figure~\ref{fig:robust:data} show the robustness of the QAE with respect to different noise types (depolarization and bitflip) and noise strengths. The solid lines depict the average energy errors (with respect to the ground state energy) or fidelity with the ground state of $\text{H}_2$ (at a bond length of 0.725$\mathring{A}$) and LiH (at the bond length 1.4$\mathring{A}$).
Here, we construct the QAE[2,1,2] and QAE[4,1,4] with $L=1$ ansatz block for $\text{H}_2$ and LiH, respectively. In each ansatz, the rotation gate $R$ is the parameterized rotation Pauli Y gate and the two-qubit entangling gates are controlled-Z placed circularly with the indices of the qubits. We denote this ansatz as RY\_CZ in our plots. Furthermore, we also compare our QAE's design with the implementation of QAE in previous research~\cite{bondarenko:2020:prl:QAE,apache:2020:QAE}, considering the optimization of all possible values of each unitary in the quantum neural network representation of QAE (see Section~\ref{sec:1A:design}). We denote this implementation as QNN, using the same optimization setting to optimize RY\_CZ.
For each random initialization, we train both structures with random 200 pairs of noisy data for 100 training epochs and test with 1000 samples of the noisy data. The predicted energy error and predicted fidelity are averaged over these 1000 samples. Furthermore, we repeat the random initialization ten times and take the average value to display in the plots.
In the fidelity plots, we also plot the average proportion of ``very noisy data" in the training data set (solid black lines). Here, we define ``very noisy data" as data with a fidelity with the ground state of less than 0.2.

Figure~\ref{fig:robust:data} implies that our method can perform well even with relatively high noise levels in the data, including a significant proportion of noisy data in the training dataset. Moreover, it outperforms the previous QNN implementation by reducing the number of training parameters for easier optimization.

\section{Denoising for VQE of Transverse Field Ising Models}
We present denoising experiments on the VQE algorithm to find the ground states of the one-dimensional (1D) Transverse Field Ising Model (TFIM).
Here, the Hamiltonian of the 1D TFIM is defined as
\begin{align}
H = - \sum_{i} J_{i}\sigma^z_i \sigma^z_{i+1} - \sum_i g_i\sigma^x_i,    
\end{align}
where $i$ refer to sites on the line. The model assumes spin-$\frac{1}{2}$ particles with Pauli matrices $\sigma^x_i$ and $\sigma^z_i$ acting on site $i$. $J_{i}$ are constants with dimensions of energy and $g_i$ are coupling parameters that determine the relative strength between the external transverse field and the nearest neighbor interactions.

In our numerical simulation, we consider uniform interaction $J_{i}=1$ and uniform onsite potential $g_i=g > 0$ with open boundary condition (i.e., $H=-\sum_{i=1}^{N-1}\sigma^z_i\sigma^z_{i+1} - g\sum_{i=1}^N\sigma^x_i$ for $N$ spins).
This model admits two phases (ordered and disordered phases), depending on whether the ground state breaks or preserves the $\prod_j\sigma^x_j$ spin-flip symmetry. 
When $g=1$, the system undergoes a quantum phase transition.

\begin{figure}
\centering
		\includegraphics[width=17cm]{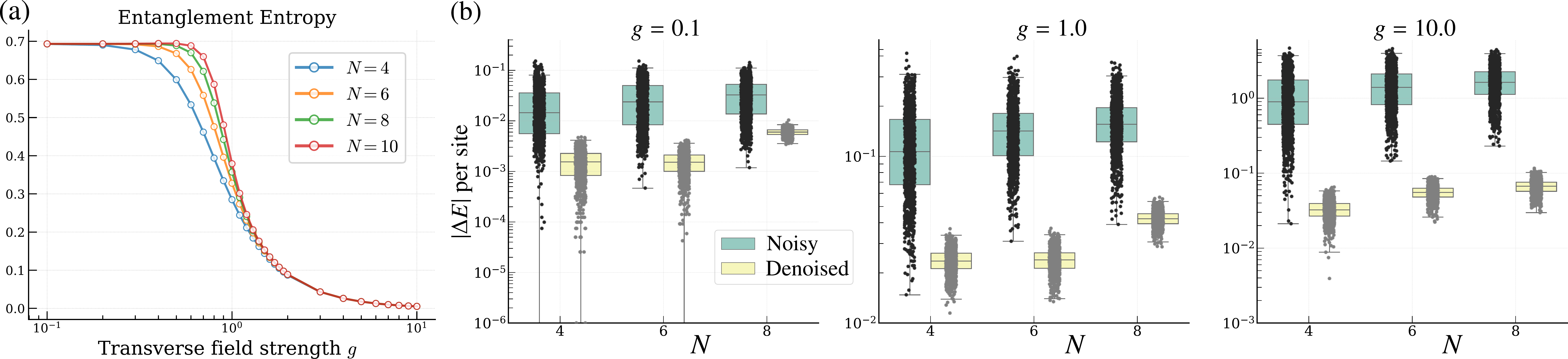}
		\protect\caption{(a) Entanglement entropy of the ground state at different values of transverse field strengths $g$ in the one-dimensional (1D) Transverse Field Ising Model (TFIM) with open boundary condition. (b) Box and jitter plots displaying the distribution of the energy error $|\Delta E|$ (the difference between estimated energy and ground state energy) over 1000 test data of 1D TFIM model at $N=4,6,8$ spins. The interquartile range is contained within the box, and the 5th and 95th percentiles are marked by the whiskers. The median is the line across the box.}\label{fig:Ising:test}
\end{figure}

To understand the entanglement behavior of 1D TFIM, we compute the entanglement entropy between the subsystem included $N/2$ spins at the left TFIM and the remained $N/2$ spins at the right.
Figure~\ref{fig:Ising:test}~(a) shows these entanglement entropy values via the variation of the transverse field strength $g$.
To demonstrate that our method can work for non-trivial states with more complexity, we apply our QAE to denoise the output of VQE to find the ground state at $g=0.1, 1.0,$ and $10.0$.
Here, the noisy data are generated from the VQE circuit at early stopping optimization with $4\times N$ SPSA iterations.
As similar with other experiments, we construct the QAE[$N,1,N$] with $L=3$ ansatz blocks where the rotation gate is the parameterized rotation Pauli Y gate $R_Y(\theta) = e^{-iY\theta/2}$ and the non-parameterized two-qubit entangling gates are controlled-Z.
We train the QAE with only 100 pairs of noisy VQE data and test with 1000 samples of noisy VQE data.

Figure~\ref{fig:Ising:test}~(b) demonstrates the
distribution of the energy error $|\Delta E|$ (the difference between estimated energy and ground state energy) over 1000 test data of 1D TFIM model at $N=4,6,8$ spins. 
While increasing $N$ can make the problem harder, our method can reduce the error by about an order of magnitude for all values of $N$ and $g$.

\begin{figure}
\centering
		\includegraphics[width=17.5cm]{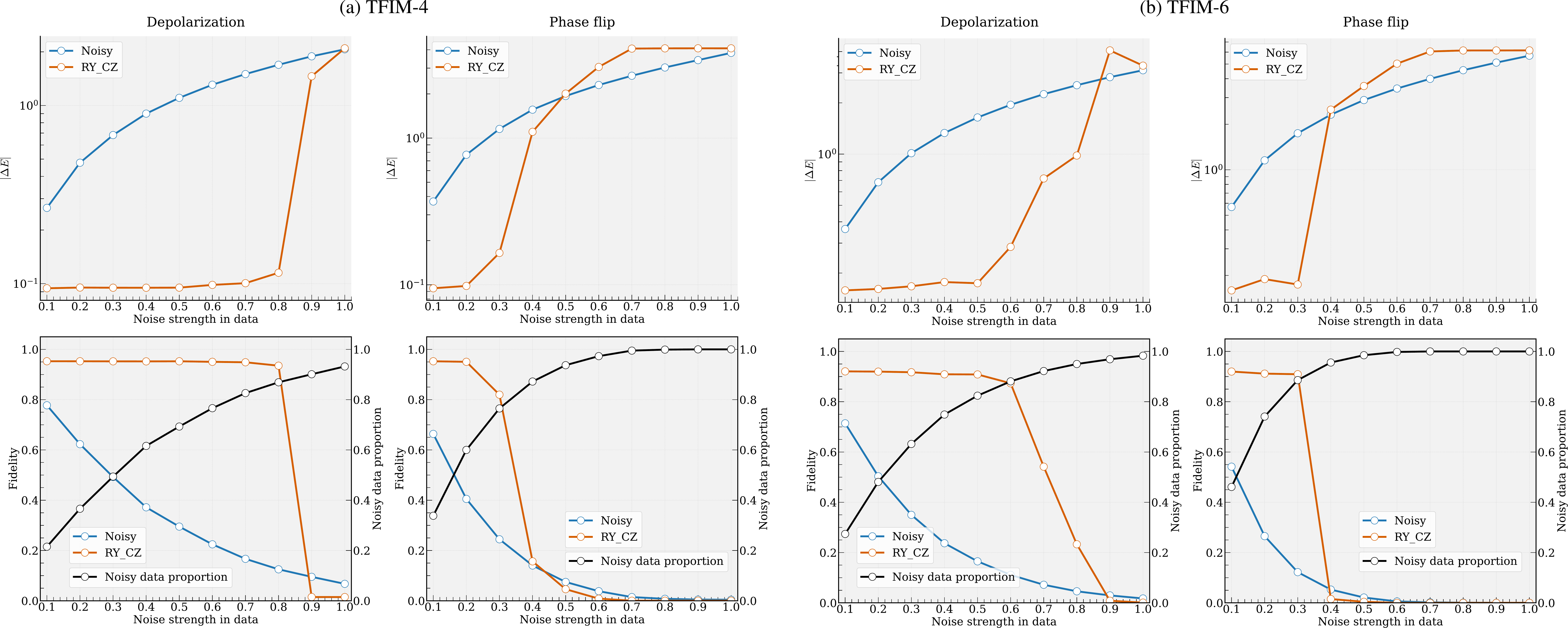}
		\protect\caption{The robustness of the QAE concerning different noise types (depolarization and phase flip) and noise strengths applied to the groundstate of (a) TFIM-4 and (b) TFIM-6 at the transverse field length $g=1.0$. The solid lines represent the average energy errors or fidelity with the ground state. The QAE[N, 1, N] (N=4,6) is constructed with an ansatz block, where the rotation gates RY are applied to all qubits, and the two-qubit controlled-Z gates are placed circularly with the indices of the qubits (denoted as RY\_CZ). 
We train both structures with 200 random pairs of noisy data for 100 training epochs and test with 1000 samples of the noisy data. The predicted energy error and predicted fidelity are averaged over these 1000 samples. In the fidelity plots, solid black lines represent the average proportion of ``very noisy data" in the training data set. Here, ``very noisy data" is defined as data with fidelity to the ground state of less than 0.2.}~\label{fig:robust:Ising:data}
\end{figure}

We further investigate the robustness of the QAE with respect to different noise types (depolarization and phase flip) and noise strengths applied to the groundstate of (a) TFIM-4 and (b) TFIM-6 at the transverse field length $g=1.0$ (Fig.~\ref{fig:robust:Ising:data}). 
The depolarizing channel is implemented by applying one of the Pauli matrices $\{I, X, Y, Z\}$ to each qubit with respective probabilities $\{1-3p_d/4, p_d/4, p_d/4, p_d/4\}$, and the phase flip channel is implemented by applying $\{I, Z\}$ to each qubit with respective probabilities $\{1-p_{\textrm{phase}}, p_{\textrm{phase}}\}$.
we vary the noise strength in the data with the noise strength $p_d, p_{\textrm{phase}} \in \{0.1, 0.2, 0.3, 0.4, 0.5, 0.6, 0.7, 0.8, 0.9, 1.0\}$. 
The solid lines depict the average energy errors (with respect to the ground state energy) or fidelity with the ground state of.
Here, we construct the QAE[4,1,4] and QAE[6,1,6] with $L=1$ ansatz block for TFIM-4 and TFIM-6, respectively. In each ansatz, the rotation gate $R$ is the parameterized rotation Pauli Y gate and the two-qubit entangling gates are controlled-Z placed circularly with the indices of the qubits. We denote this ansatz as RY\_CZ in our plots. We train both structures with random 200 pairs of noisy data for 100 training epochs and test with 1000 samples of the noisy data. The predicted energy error and predicted fidelity are averaged over these 1000 samples.
In the fidelity plots, we also plot the average proportion of ``very noisy data" in the training data set (solid black lines). Here, we define ``very noisy data" as data with a fidelity with the ground state of less than 0.2.

As similar to Fig.~\ref{fig:robust:data}, Figure~\ref{fig:robust:Ising:data} implies that our method can perform well even with relatively high noise levels in the data, including a significant proportion of noisy data in the training dataset.

\section{Design the ansatz for QAE}

The ansatz structure provided in our algorithm relies on hardware-efficient ansatzes.
\begin{align}\label{eqn:ansatz}
    U_j^i(\bth) = \prod_{l=1}^L \left[R^{(l)}_{\text{loc}}(\bth_l)V_{\text{ent}}\right] R^{(L+1)}_{\text{loc}}(\bth_{L+1})
\end{align}
Here, each unitary $U_j^i$ consists of $L$ blocks and final single qubit rotation on every qubit. Each block comprises of local single qubit rotations $R^{(l)}_{\text{loc}}(\bth_l)=\otimes_k R(\theta_{k,l})$ as well as two-qubit entangling gates. 
In our experiments, we use the ansatz RY\_CZ, where the single-qubit rotation gate $R$ is the parametric rotation gate $R_Y(\theta) = e^{-iY\theta/2}$ of Pauli Y matrix, and the entangler $V_{\text{ent}}$ is composed of controlled-phase gates $CZ(k,k+1)$ placed in circular with indexes $k$ of qubits.

We can provide a recipe for selecting the ansatz. In our study, for instance, we considered several ansatz structures, trained them with a small amount of data, and selected the ansatz with the best performance. We conducted a numerical investigation of several common ansatz types to support the selection of the RY\_CZ ansatz in our study.
In our numerical experiments, we examined six ansatz patterns: RY\_CZ, RY\_CX, RYZ\_CZ, RY\_CAN, RYZ\_CAN, and RZYY\_CAN. Here, the parametric rotation gate consists of Pauli Y and Pauli Z matrices in local single-qubit rotations, and the entangler $V_{\text{ent}}$ is either a CX, CZ, or two-qubit canonical (CAN) gate.
The two-qubit canonical gate is defined as the composition of three two-qubit gates
\begin{align}
    \text{CAN} = \text{RXX}(\pi)\text{RYY}(\pi)\text{RZZ}(\pi) = e^{-i\frac{\pi}{2}X\otimes X}e^{-i\frac{\pi}{2}Y\otimes Y}e^{-i\frac{\pi}{2}Z\otimes Z}.
\end{align}
Notably, the last pattern RZYY\_CAN can be employed to implement any arbitrary two-qubit unitary operation.

\begin{figure}
\centering
	\includegraphics[width=17cm]{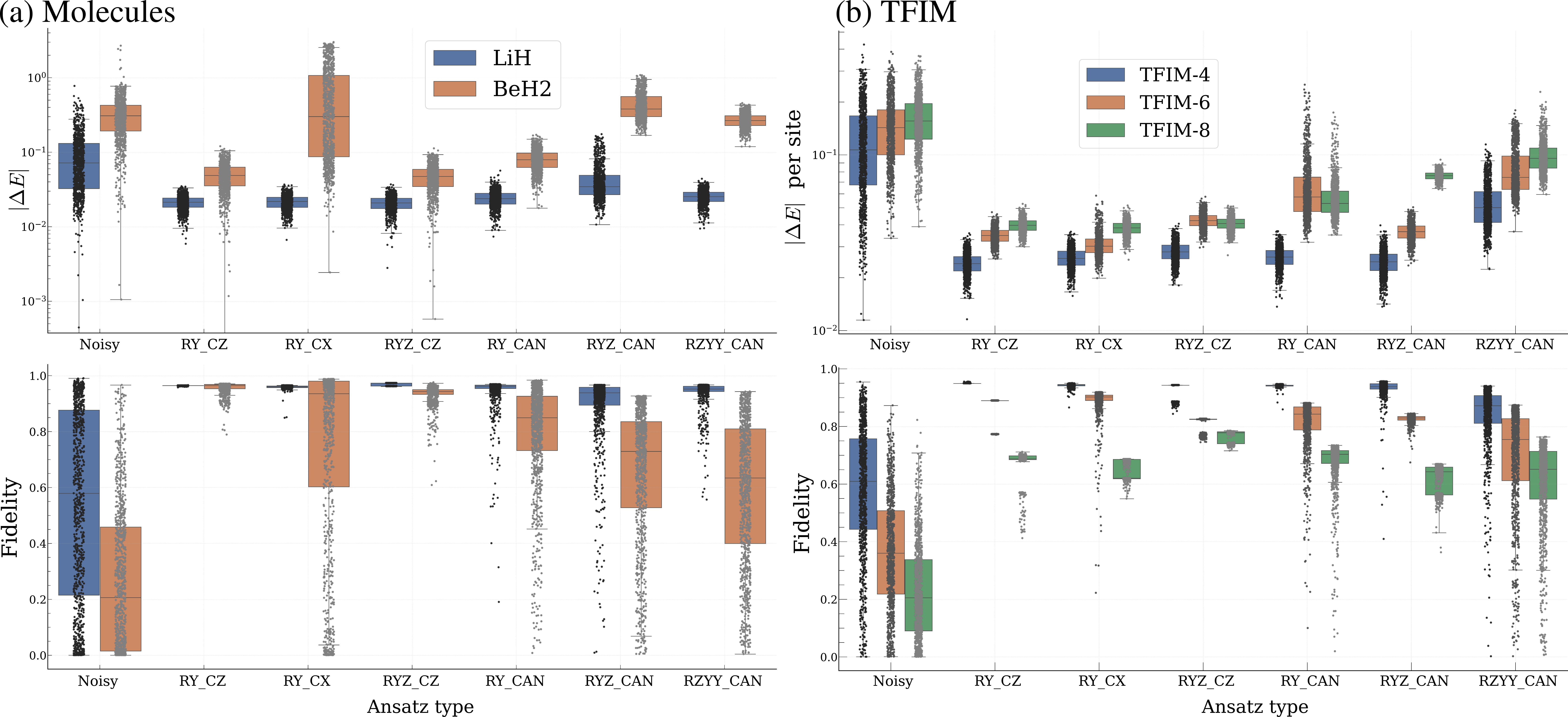}
		\protect\caption{Box and jitter plots display the distribution of the energy error $|\Delta E|$ and the fidelity with the ground state over 1000 test data samples for (a) LiH (at a bond length of 1.595$\mathring{A}$, with 100 SPSA iterations in the VQE routine) and $\text{BeH}_2$ (at a bond length of 1.6$\mathring{A}$, with 500 SPSA iterations in the VQE routine) molecules, (b)  1D TFIM model at $g=1.0$ and N=4, 6, 8 qubits with $10\times N$ SPSA iterations in the VQE routine, considering different QAE ansatz types. The interquartile range is contained within the box, and the 5th and 95th percentiles are marked by the whiskers. The median is the line across the box.}\label{fig:ansatz:select}
\end{figure}

We use these ansatz patterns in the denoising tasks for LiH (at a bond length of 1.595$\mathring{A}$ and 100 VQE iterations), $\text{BeH}_2$ (at a bond length of 1.6$\mathring{A}$ and 500 VQE iterations), and 1D TFIM model at $g=1.0$ and N = 4, 6, and 8 qubits with $10\times N$ VQE iterations.
Figure~\ref{fig:ansatz:select} depicts the distribution of the energy error $|\Delta E|$ and fidelity with the ground state over 1000 test data points to demonstrate that RY\_CZ is the most suitable option for our ansatz. Furthermore, for molecules LiH and $\text{BeH}_2$, while the CX gate is used in the VQE ansatz, it cannot produce good results in denoising. Therefore, the ansatz of QAE is not necessary the same with the VQE or universal.

\begin{figure}
\centering
	\includegraphics[width=17.5cm]{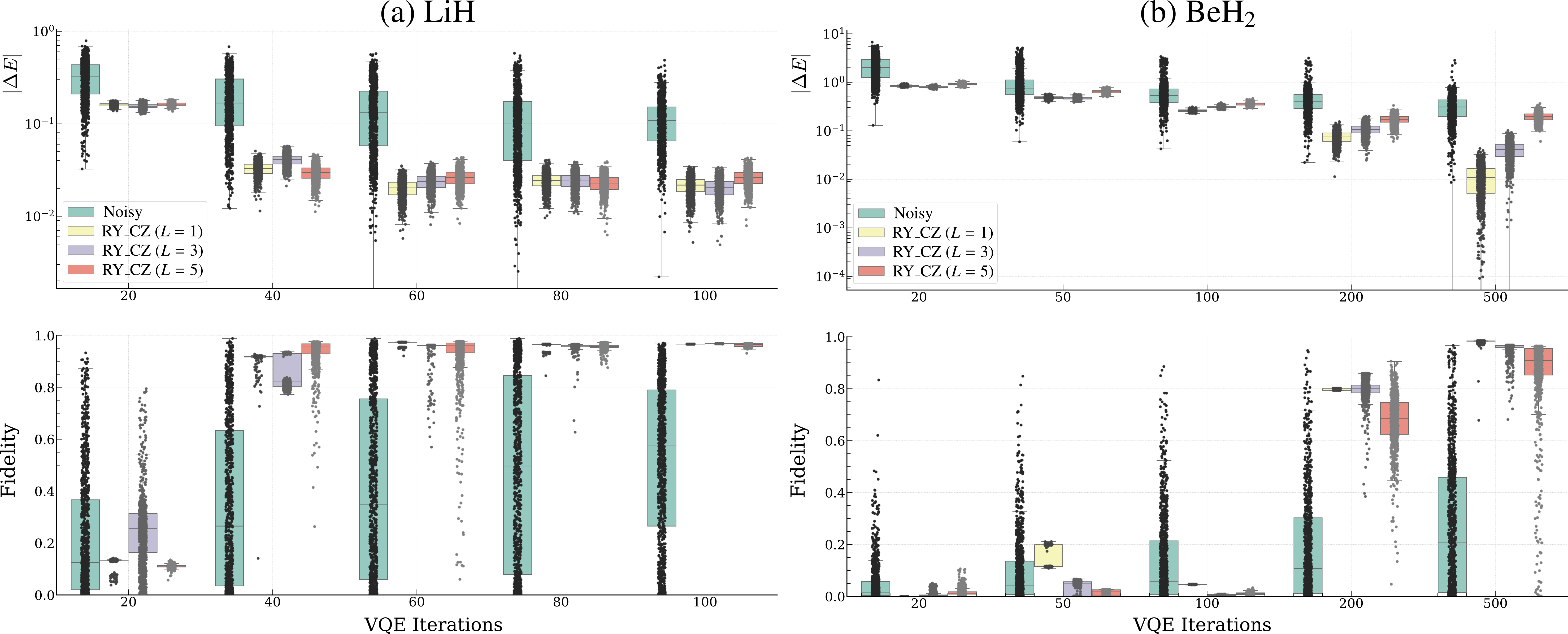}
		\protect\caption{Box and jitter plots display the distribution of the energy error $|\Delta E|$ and the fidelity with the ground state over 1000 test data samples for (a) LiH (at a bond length of 1.595$\mathring{A}$) and (b) $\text{BeH}_2$ (at a bond length of 1.6$\mathring{A}$) molecules. These plots illustrate the performance under different number $L$ of ansatz blocks and different number of SPSA iterations in the VQE routine to create original training data, where shorter iterations correspond to noisier data.
The interquartile range is contained within the box, and the 5th and 95th percentiles are marked by the whiskers. The median is represented by the line across the box.}\label{fig:iters:block:mol}
\end{figure}

\begin{figure}
\centering
	\includegraphics[width=17.5cm]{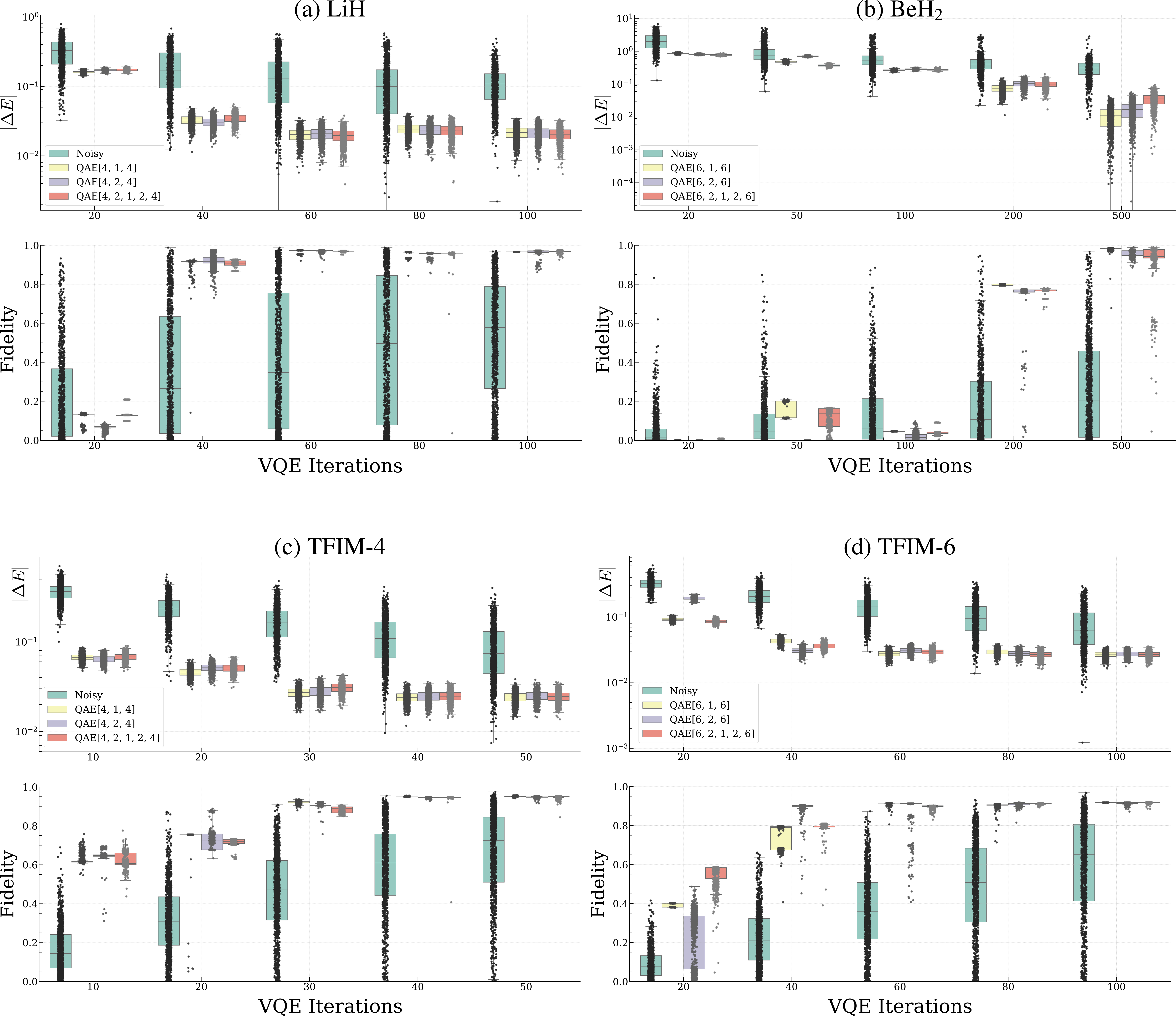}
		\protect\caption{Box and jitter plots display the distribution of the energy error $|\Delta E|$ and the fidelity with the ground state over 1000 test data samples for (a) LiH (at a bond length of 1.595$\mathring{A}$) and (b) $\text{BeH}_2$ (at a bond length of 1.6$\mathring{A}$) molecules, and 1D TFIM model at $g=1.0$ and (c) N=4 (TFIM-4) and (d) N=6 qubits (TFIM-6). These plots illustrate the performance under different QAE topologies and SPSA iterations in the VQE routine to create original training data, where shorter iterations correspond to noisier data.
The interquartile range is contained within the box, and the 5th and 95th percentiles are marked by the whiskers. The median is represented by the line across the box.}\label{fig:iters:topo}
\end{figure}

We further investigate the denoising performance when increasing the number $L$ of ansatz blocks described in Eq.~\eqref{eqn:ansatz}. Figure~\ref{fig:iters:block:mol} illustrates the performance of LiH and $\text{BeH}_2$ molecules under different number $L$ of ansatz blocks and different number of VQE iterations in the original training data, where shorter iterations correspond to noisier data. 
We can also investigate the denoising performance when consider different configurations of QAE. Figure~\ref{fig:iters:topo} illustrates the performance under different QAE's topology and different number of SPSA iterations in the VQE routine to create the original training data, where shorter iterations correspond to noisier data. While increasing the complexity of QAE's topology and the number of ansatz blocks can increase the expressivity of QAE, it makes the training harder with a large number of parameters. In our tasks, we set the topology of QAE as $[N, 1, N]$ with one bottleneck node, and $L=1$ or $L=3$ as it is sufficient to see the effect of the denoising performance. However, we must improve the optimizer or choose the initialization carefully if we want to deal with a large number $L$ or more complex QAE's topology.

% \begin{figure}
% \centering
% 	\includegraphics[width=17.5cm]{iter_TFIM_multi_L_test_1000_train_200_0_200_batch_50_ep_100.pdf}
% 		\protect\caption{\diff{Box and jitter plots display the distribution of the energy error $|\Delta E|$ and the fidelity with the ground state over 1000 test data samples for TFIM. These plots illustrate the performance under different number $L$ of ansatz blocks and different number qubits $N$.
% The interquartile range is contained within the box, and the 5th and 95th percentiles are marked by the whiskers. The median is represented by the line across the box.}\label{fig:iters:block:mol}}
% \end{figure}

\clearpage
\section{Barren plateaus in training QAE}

In principle, the circuits employed in our study (RY\_CZ) and the global function cannot entirely eliminate barren plateaus, especially when scaling up the problem.
However, our research has identified practical techniques to mitigate the barren plateau issue. Initially, we train the QAE using a small set of training data to establish a relatively good starting point for the parameters. Subsequently, during the training process, we employ a combination of SPSA with AMSGrad in mini-batch training, incorporating stochastic parameter updates to enhance the variation in the approximated gradient.

\begin{figure}
\centering
		\includegraphics[width=12cm]{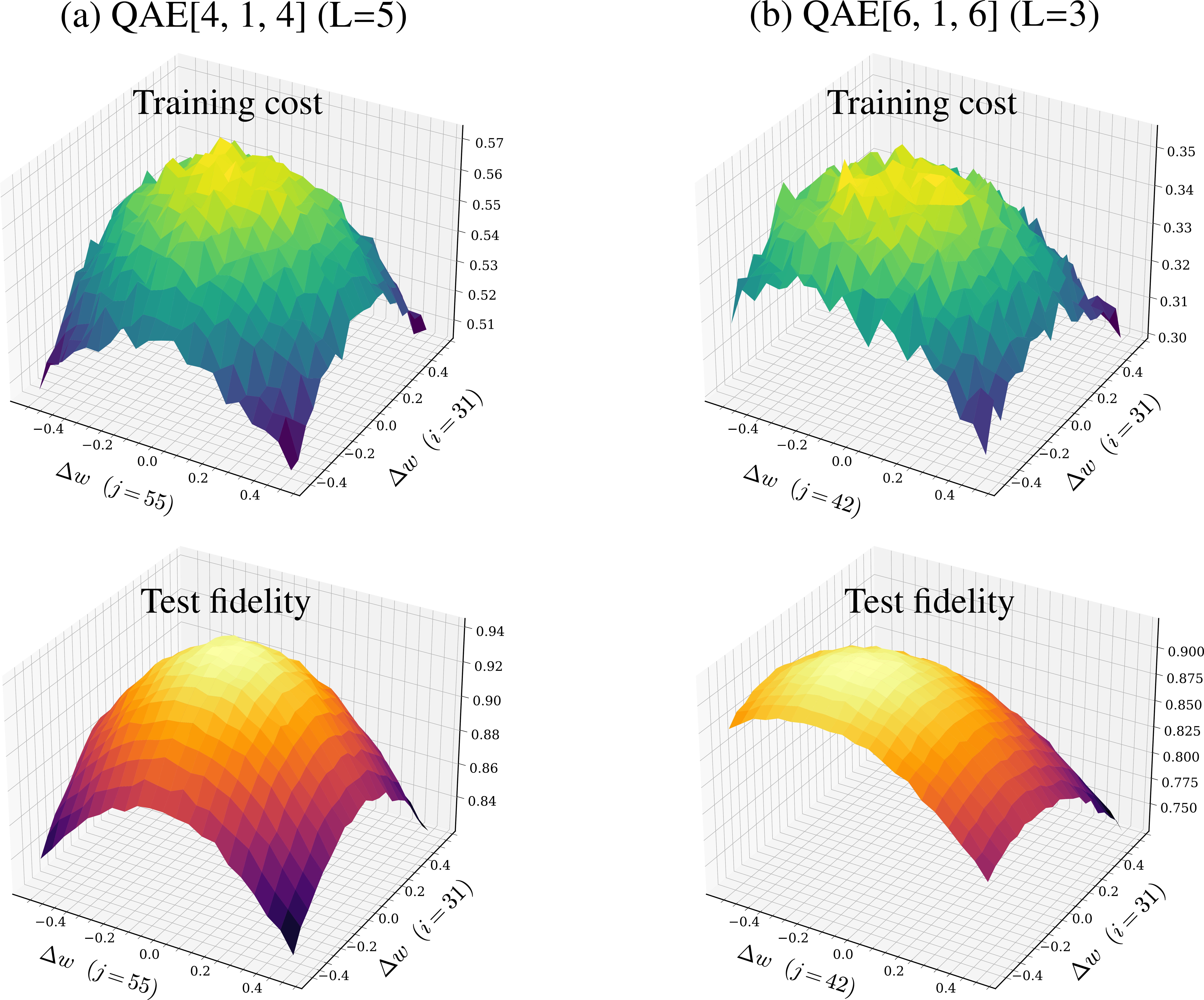}
		\protect\caption{The training cost and test fidelity landscapes of QAE[4, 1, 4] and QAE[6, 1, 6] with $L=5$ and $L=3$ ansatz blocks, respectively. We maintain other parameters fixed but vary $\theta_i \in \Omega_i$ and $\theta_j \in \Omega_j$ to plot the training cost and test fidelity. Here, $\Omega_k = [\theta^{t}_k-\Delta w, \theta^{t}_k + \Delta w]$ (with $\Delta w = 0.5$) represents the interval around the trained value $\theta^t_k$ of the parameter $\theta_k$. }\label{fig:Ising:landscape}
\end{figure}

We expect that training with noisy data can alleviate the difficulties encountered when scaling up. For instance, in Fig.~\ref{fig:Ising:landscape}, we depict the training cost and test fidelity landscapes of QAE[4, 1, 4] and QAE[6, 1, 6] with $L=5$ and $L=3$ ansatz blocks in each unitary of the dissipative QAE, respectively. Here, we initially train the QAE for the transverse field Ising model and select each parameter $\theta_k$ to vary around the trained value $\theta^{t}_k$ as $\Omega_k = [\theta^{t}_k-\Delta w, \theta^{t}_k + \Delta w]$ ($\Delta w = 0.5$), while keeping other parameters fixed. We choose two indices, $i$ and $j$, such that the training cost varies the most in $\Omega_i$ and $\Omega_j$. Finally, we maintain the other parameters fixed but vary $\theta_i \in \Omega_i$ and $\theta_j \in \Omega_j$ to plot the training cost and test fidelity in Fig.~\ref{fig:Ising:landscape}. Even with a relatively large number of parameters (78 and 76), we can observe that the training landscape does not suffer from the barren plateau at the current scale. However, we acknowledge that this does not imply the absence of the barren plateau in larger systems.
Exploring the design of the QAE with training mechanisms and data properties that can circumvent the barren plateau is an intriguing avenue for future research. 

\section{The denoising mechanism}
In the main text, we provided the discussion on the understanding of the mechanism behind QAE in denoising. We further present additional results in this section.

We investigate the fidelities between the outputs of subsystems of the QAE and the corresponding parts in the target ground state. To illustrate this, let's consider the QAE[6,1,6]. First, the entire QAE structure [6,1,6] (not just a subspace) is trained. Then, we test the trained QAE with noisy data, which involves computing the fidelity between the output of QAE[6,1,6] and the true ground state across 1000 noisy data samples.

In this testing phase, we introduce a unique approach by employing inactive neurons. Inactive neurons refer to neurons for which connections have been removed in the trained QAE. For instance, if we set the 6th neuron in the last layer as inactive, we essentially have a QAE[6,1,5] with a subset of the trained parameters from QAE[6,1,6]. The output of this QAE comprises five qubits, which is then compared with the state of the first five qubits in the target ground state. We apply a similar procedure when we set the 5th and 6th neurons in the last layer as inactive, leading to a QAE[6,1,4] (and continue with QAE[6,1,3], QAE[6,1,2], and QAE[6,1,1] by inactivating corresponding neurons sets), and compare the output of this QAE with the first four qubits (or the first three qubits, the first two qubits, and the first qubit, respectively) in the target ground state.

\begin{figure}
\centering
		\includegraphics[width=12cm]{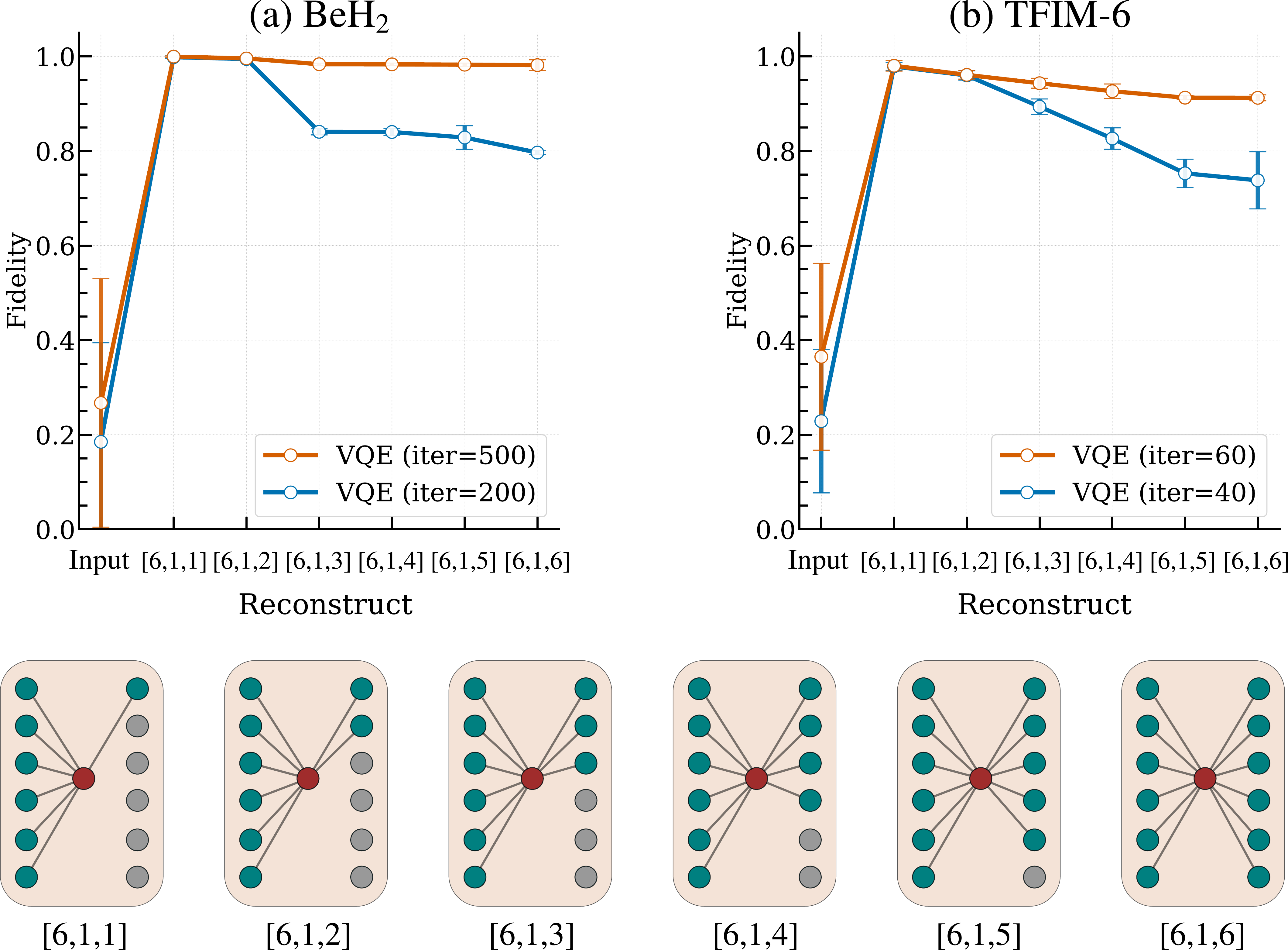}
		\protect\caption{Fidelities between the outputs of subsystems of the QAE[6,1,6] and the corresponding parts in the ground state of BeH2 and TFIM-6. Here, the noisy data is generated from VQE circuits at different SPSA iterations (shorter iterations mean higher noise levels). The gray node represents inactive neurons, where subsystems  [6,1,1], [6,1,2], [6,1,3], [6,1,4], and [6,1,5] correspond the subsystems with the first qubit, the first two qubits, the first three qubits, the first four qubits, and the first five qubits of the ground state, respectively. As the noise level increases, noisy components leak out from the bottleneck layer, preventing the decoder from reconstructing the ground state. Solid lines and error bars describe the average values and standard deviations computed over 1000 test samples.
		\label{fig:layers}}
\end{figure}

This investigation provides insights into how the QAE processes noisy data with varying noise levels. For instance, in Fig.~\ref{fig:layers}(a), we consider noisy data generated from VQE circuits of BeH2 (at a bond length of 1.6$\mathring{A}$) after 500 (low noise) and 200 (high noise) SPSA iterations. When the input state is processed through the QAE, a portion of the target state is reconstructed in the subsystem's output.

At a low noise level, for instance, when VQE output states are generated after 500 SPSA iterations, for subsystems [6,1,1], [6,1,2], [6,1,3], [6,1,4], and [6,1,5], the QAE can reconstruct the first qubit, the first two qubits, the first three qubits, the first four qubits, the first five qubits, and eventually all qubits of the target state (represented by the orange lines) with high fidelity. However, if we train the QAE with high noise-level VQE data where the VQE output states (after 200 SPSA iterations) have low overlap with the ground state, irrelevant information associated with the ground state escapes at the bottleneck, hindering the decoder from reconstructing the ground state (represented by the blue lines).
The similar observation has been further confirmed in the  experiment with TFIM-6, as shown in Fig.~\ref{fig:layers}(b). We consider noisy data generated from VQE circuits of TFIM-6 (at $g=1.0$) after 50 (low noise) and 40 (high noise) SPSA iterations.

%\bibliography{main.bib}
\providecommand{\noopsort}[1]{}\providecommand{\singleletter}[1]{#1}%